\documentclass[journal,10pt,twoside,twocolumn]{IEEEtran}

\usepackage{cite}
\usepackage[T1]{fontenc}
\usepackage{graphicx}
\usepackage{amssymb}
\usepackage{amsmath}
\usepackage{microtype}
\usepackage{hyperref}
\usepackage{url}
\usepackage{balance}
\usepackage{fancyref}
\usepackage{hyperref}
\newcommand{\cshere}[1]{\begin{center}\textcolor{red}{\bf--- CS stopped his edits here ---}\end{center}}

\renewcommand{\eqref}[1]{(\ref{#1})}
\newcommand{\secref}[1]{\mbox{Section~\ref{#1}}}
\newcommand{\figref}[1]{\mbox{Figure~\ref{#1}}}
\newcommand{\C}{\ensuremath{\mathbb{C}}}
\newcommand{\R}{\ensuremath{\mathbb{R}}}

\newcommand{\setD}{\ensuremath{\mathcal{D}}}

\newcommand{\setF}{\ensuremath{\mathcal{F}}}

\newcommand{\setW}{\ensuremath{\mathcal{W}}}

\newcommand{\bma}{\ensuremath{\mathbf{a}}}

\newcommand{\bmn}{\ensuremath{\mathbf{n}}}

\newcommand{\bms}{\ensuremath{\mathbf{s}}}

\newcommand{\bmw}{\ensuremath{\mathbf{w}}}
\newcommand{\bmx}{\ensuremath{\mathbf{x}}}
\newcommand{\bmy}{\ensuremath{\mathbf{y}}}

\newcommand{\bmphi}{\ensuremath{\boldsymbol\phi}}

\newcommand{\bA}{\ensuremath{\mathbf{A}}}

\newcommand{\bF}{\ensuremath{\mathbf{F}}}

\newcommand{\bI}{\ensuremath{\mathbf{I}}}

\newcommand{\bR}{\ensuremath{\mathbf{R}}}

\newcommand{\bW}{\ensuremath{\mathbf{W}}}

\newcommand{\bPsi}{\ensuremath{\mathbf{\Psi}}}

\newcommand{\bTheta}{\ensuremath{\mathbf{\Theta}}}

\newcommand{\bPhi}{\ensuremath{\mathbf{\Phi}}}

\newcommand{\matb}{\left( \begin{matrix*}[r] }
\newcommand{\mate}{\end{matrix*}\right)}

\usepackage{color}
\newcommand{\revision}[1]{#1}

\begin{document}

\title{Non-Uniform Wavelet Sampling for  \\ RF Analog-to-Information Conversion}
\author{Micha\"el Pelissier and Christoph Studer\thanks{M. Pelissier was a visiting researcher at the School of Electrical and Computer Engineering (ECE), Cornell University, Ithaca, NY, from CEA \revision{Univ. Grenoble Alpes, CEA, LETI, F-38000 Grenoble }, France (e-mail: \url{ michael.pelissier@cea.fr}).}\thanks{C.~Studer is with the School of ECE, Cornell University, Ithaca, NY (e-mail: \url{studer@cornell.edu}). Website: \url{vip.ece.cornell.edu}}}

%\thanks{Digital Object Identifier XXX-XXX-XXX} 

\maketitle

% ================================================================================
% ================================================================================
% ================================================================================
%\vspace{-1.1cm}
\begin{abstract}
Feature extraction, such as spectral occupancy, interferer energy and type, or direction-of-arrival, from wideband radio-frequency~(RF) signals finds use in a growing number of applications as it enhances RF transceivers with cognitive abilities and enables parameter tuning of traditional RF chains. In power and cost limited applications, e.g., for sensor nodes in the Internet of Things, wideband RF feature extraction  with conventional, Nyquist-rate analog-to-digital converters is infeasible. However, the structure of many RF features (such as signal sparsity) enables the use of compressive sensing (CS) techniques that acquire such signals at sub-Nyquist rates. \revision{While such CS-based analog-to-information (A2I) converters have the potential to enable low-cost and energy-efficient wideband RF sensing, they suffer from a variety of real-world limitations, such as noise folding, low sensitivity, aliasing, and limited flexibility. }

\revision{This paper proposes a novel CS-based A2I architecture called non-uniform wavelet sampling (NUWS). Our solution extracts a carefully-selected subset of wavelet coefficients directly in the RF domain, which mitigates the main issues of existing A2I converter architectures. For multi-band RF signals, we propose a specialized variant called non-uniform wavelet bandpass sampling (NUWBS), which further improves sensitivity and reduces hardware complexity by leveraging the multi-band signal structure. We use simulations to demonstrate that NUWBS approaches the theoretical performance limits of $\bf\boldsymbol\ell_1$-norm-based sparse signal recovery. We investigate hardware-design aspects and show ASIC measurement results for the wavelet generation stage, which highlight the efficacy of NUWBS for a broad range of RF feature extraction tasks in cost- and power-limited applications.}
\end{abstract}


% ================================================================================
% ================================================================================
% ================================================================================

\section{Introduction}
\label{sec:introduction}
\IEEEPARstart{F}{or} nearly a century, the cornerstone of digital signal processing has been the Shannon--Nyquist--Whittaker sampling theorem~\cite{shannon1949communication}. \revision{This result states that signals of finite energy and bandwidth are perfectly represented by a set of uniformly-spaced samples at a rate higher than twice the maximal frequency.} It is, however, well-known that signals with  certain structure can be sampled well-below the Nyquist rate. For example, Landau established in 1967 that multi-band signals occupying $N$ non-contiguous frequency bands of bandwidth~$B$ can be represented using an average sampling rate no lower than twice the sum of the bandwidths (i.e.,  $2NB$)~\cite{landau1967sampling}. \revision{In~2006, Landau's concept has been generalized by Cand\`es, Donoho, Romberg, and Tao in \cite{candes2006robust,donoho2006compressed} to sparse signals, i.e., signals that have only a few nonzero entries in a given transform basis, e.g., the discrete Fourier transform~(DFT). These  results are known as \emph{compressive sensing}~(CS) and find potential use in a broad range of sampling-critical applications~\cite{candes2008introduction}.}

In essence, CS fuses sampling and compression: instead of sampling signals at the Nyquist rate followed by conventional data compression, CS acquires ``just enough'' compressive measurements that guarantee the recovery of the signal of interest. Signal recovery then exploits the concept of sparsity, a structure that is present in most natural and man-made signals.
\revision{CS has the potential to acquire signals with sampling rates well-below the Nyquist frequency, which may lead to significant reductions in the sampling costs and/or power consumption, or enable an increase the bandwidth of signal acquisition beyond the physical limits of analog-to-digital converters (ADCs)~\cite{murman2016survey}.} As a consequence, CS is commonly believed to be a panacea for wideband radio-frequency~(RF) spectrum awareness applications \cite{sharma2016application,sun2013wideband,baccour2012radio}.

\subsection{Challenges of Wideband Spectrum Awareness} % 
In RF communication, there is a growing need in providing radio transceivers with cognitive abilities that enable awareness and adaptability to the spectrum environment \cite{baccour2012radio}. 
The main goals of suitable methods are to capture a variety of  RF parameters (or features) to dynamically allocate spectral resources \cite{khan2015cognitive} and/or to tune  traditional RF-chain circuitry with optimal parameter settings in real-time, e.g.,  \revision{to cancel out strong interferers using a tunable notch filter \cite{yazicigil201519,adams2016amixer}.}  
The RF features to be acquired for these tasks are mainly related to \emph{wideband spectrum sensing}~\cite{sun2013wideband} and include the estimation of frequency occupancy, signal energy, energy variation, signal-to-noise-ratio, direction-of-arrival, etc.~\cite{sharma2016application,hayashi2013auser}. 

For most wideband spectrum sensing tasks, one is typically interested in acquiring large bandwidths (e.g., several GHz) with a high dynamic range (e.g., $80$\,dB or more). However, achieving such specifications with a single analog-to-digital converter~(ADC) is an elusive goal with current semiconductor technology~\cite{murman2016survey}. A practicable solution is to scan the entire bandwidth in sequential manner. From a hardware perspective, this approach relies on traditional RF receivers as put forward by Armstrong in 1921~\cite{armstrong1921new}. The idea is to mix the incoming RF signal with a complex sinusoid (whose frequency can be tuned) either to a lower (and fixed) frequency or directly to baseband. The signal is then sampled with an ADC operating at lower bandwidth. While such an approach enjoys widespread use---mainly due to its excellent spectral selectivity, sensitivity, and dynamic range---the associated hardware requirements (for wideband tunable oscillators and highly-selective filters) and sweeping time may not meet real-world application constraints~\cite{yazicigil201519}. 
This aspect is particularly important for the Internet of Things (IoT), in which devices must adhere to stringent power and cost constraints, while operating in a multi-standard environment (e.g., containing signals from 3GPP NB-IoT, IEEE 802.15.4g/15.4k/11.ah, SigFox, and LoRa). Hence,  there is a pressing need for RF feature extraction methods that minimize the power and system costs, while offering flexibility to a variety of environments and standards.

\subsection{Analog-to-Information (A2I) Conversion}
A promising solution for such  wideband spectrum sensing applications is to use CS-based analog-to-information (A2I) converters that leverage spectrum sparsity  \cite{wakin2012nonuniform,bellasi2013vlsi,sun2013wideband,yazicigil2015wideband}.
%\cite{}\cite{}
Indeed, one of the main advantages of CS is that it enables the acquisition of larger bandwidths with relaxed sampling-rate requirements, \revision{thus enabling less expensive, faster, and potentially more energy-efficient solutions.}  While a large number of CS-based A2I converters have been proposed for spectrum sensing tasks \cite{sun2013wideband,bellasi2013vlsi,wakin2012nonuniform,chen2011sub,yazicigil201519}, the generally-poor noise sensitivity of traditional CS methods~\cite{davenport2012pros,arias2011noise} and the excessive complexity of the recovery stage~\cite{maechler2012vlsi,bellasi2013vlsi} prevents their straightforward use in low-power, cost-sensitive, and latency-critical applications, which are typical for the  IoT.

Fortunately, for a broad range of  RF feature extraction tasks, recovery of the entire spectrum or signal may not be necessary. In fact, as it has been noted in \cite{verhelst2015analog}, only a small number of measurements may be required if one is interested in certain signal features  and not the signal itself.
This key observation is crucial for a broad range of emerging energy or cost-constrained  applications in the RF domain, such as sensors or actuators for the IoT,  wake-up radio, spectrum sensing, and radar applications~\cite{juhwan2012acompressed,tian2012cyclic,magno2014wake,sharma2016application}.
In most of these applications, the information of interest has, informally speaking, a rate far below the physical bandwidth, \revision{which is a perfect fit for CS-based A2I converters that have the potential to acquire the relevant features directly in the RF (or analog) domain at  low cost and low power.} 

\setlength{\textfloatsep}{10pt}% Remove \textfloatsep
\begin{figure}[tp]
\centering
\includegraphics[width=0.9\columnwidth]{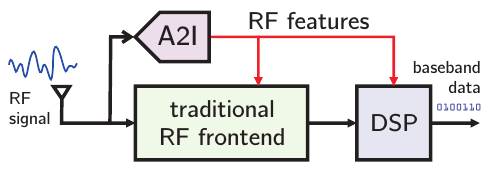}
\vspace{-0.1cm}
\caption{Overview of a cognitive radio receiver: A traditional RF front-end is enhanced with an analog-to-information (A2I) converter that extracts RF features directly from the incoming analog RF signals. The \revision{A2I converter} enables parameter tuning to reduce design margins in the RF circuitry and assists spectrum awareness tasks in the digital signal processing (DSP) stage.}
\label{fig:architecture}
\end{figure}

\figref{fig:architecture} illustrates a cognitive radio receiver that is assisted with an A2I converter specifically designed for RF feature extraction. The A2I converter bypasses conventional RF circuitry and extracts a small set of features directly from the incoming RF signals in the analog domain. The acquired features can then be used by the RF front-end and/or a subsequent digital signal processing (DSP) stage. 
\revision{Such an A2I-assisted RF front-end enables one to optimally reconfigure the key parameters of a traditional RF chain according to the spectral environment. This capability can also be used to assist traditional RF transceivers  by providing means to eliminate over-design margins through adaptation to the spectrum environment via radio link-quality estimation and interferer localization~\cite{baccour2012radio}, which is relevant in power- and cost-limited IoT applications.}

\subsection{Contributions}
This paper proposes a novel CS-based A2I converter architecture for cognitive RF receivers.  \revision{Our approach, referred to as \emph{non-uniform wavelet sampling (NUWS)}, combines wavelet preprocessing with non-uniform sampling in order to alleviate the main issues of existing A2I converter solutions, such as signal noise, aliasing, and stringent clocking constraints, which enables a broad range of RF feature extraction tasks.}
For RF multi-band signals, we propose a specialized  variant called \emph{non-uniform wavelet bandpass sampling (NUWBS)}, which combines traditional bandpass sampling with NUWS. This solution builds upon (i) wavelet sample acquisition using highly over-complete  and hardware-friendly Gabor frames or Morlet wavelets and (ii) a suitable measurement selection strategy that identifies the relevant wavelets required for RF feature extraction.
\revision{We use system simulations to demonstrate the efficacy of NUWBS and show that it approaches the theoretical phase transition of $\ell_1$-norm-based sparse signal recovery in typical  multi-band RF applications.}
We investigate hardware-implementation aspects  and validate the effective interference rejection capability of NUWBS.
\revision{We conclude by showing ASIC measurement results for the wavelet generation stage in order to highlight the practical feasibility of the wavelet generator, which is at the heart of the proposed A2I converter architecture.}

\subsection{Paper Outline}
The rest of the paper is organized as follows. 
\secref{sec:CStechniques} provides an introduction to CS and discusses existing A2I \revision{converter} architectures for sub-Nyquist RF signal acquisition.
\secref{sec:NUWS} presents our non-uniform wavelet sampling (NUWS) method and the specialized variant for multi-band signals called non-uniform wavelet bandpass sampling (NUWBS). 
\secref{sec:NUWBSdesign} discusses optimal measurement selection strategies and provides simulation results. 
\secref{sec:implementation} discusses hardware implementation aspects of NUWBS.
We conclude in \secref{sec:conclusions}.

\subsection{Notation}
\revision{Lowercase and uppercase boldface letters denote column vectors and matrices, respectively. 
For a matrix $\bA$, we represent its transpose and Hermitian transpose by $\bA^T$ and~$\bA^H$, respectively. The $M\times M$ identity matrix is $\bI_M$.
The entry on the $k$th row and  $\ell$th column of $\bA$ is $[\bA]_{k,\ell}=A_{k,\ell}$ and the $\ell$th column is $[\bA]_{:,\ell}=\bma_\ell$; the $k$th entry of the vector $\bma$ is~$[\bma]_k=a_k$. 
We write  $\bR_\Omega\bA=[\bA]_{\Omega,:}$ and $\bR_\Omega\bma_\Omega=[\bma]_\Omega$ to restrict the rows of a matrix $\bA$ and the entries of a vector $\bma$ to the index set $\Omega$, respectively.
 Continuous and discrete-time signals are denoted by $x(t)$ and $x[n]$, respectively.}
%

% ================================================================================
% ================================================================================
% ================================================================================

\section{CS Techniques for RF Signal Acquisition}
\label{sec:CStechniques}

We start by introducing the basics of CS and then review the most prominent A2I \revision{converter} architectures for RF signal acquisition, namely  non-uniform sampling (NUS) \cite{venkataramani2000perfect,venkataramani2001optimal,lazar2004perfect,mishali2009blind,wakin2012nonuniform,bellasi2013vlsi}, variable rate sub-Nyquist sampling~\cite{fleyer2010multirate,bai2012compressive,sun2013wideband,tzou2015low}, and random modulation~\cite{tropp2010beyond,pareschi2015hardware}, which includes the modulated wideband converter and Xampling~\cite{mishali2010theory,mishali2011sub,mishali2011xampling,yazicigil201519}. \revision{For each of these  architectures, we briefly discuss the pros and cons from a RF spectrum sensing and hardware design standpoint.}

\subsection{\revision{Compressive Sensing (CS) Basics}}
Let $\bmx\in\C^N$ be a discrete-time, $N$-dimensional complex-valued signal vector that we wish to acquire. We assume that the signal $\bmx$ has a so-called $K$-sparse representation $\bms\in\C^N$, i.e., the vector $\bms$ has $K$ dominant non-zero entries in a known (unitary) transform basis $\bPsi\in\C^{N\times N}$ with $\bmx=\bPsi\bms$ and $\bPsi^H\bPsi=\bI_N$. In spectrum sensing applications, one typically assumes sparsity in the DFT basis, i.e., $\bPsi=\bF^H$ is the $N$-dimensional inverse DFT matrix.  
CS acquires $M$ compressive measurements as $y_i=\langle \bmphi_i, \bms\rangle + n_i$ for $i=1,2,\ldots,M$, where $\bmphi_i\in\C^N$ are the measurement vectors and $n_i$ models measurement noise.
The CS measurement process can be written in compact matrix-vector form as follows:
\begin{align} \label{eq:acquisition}
\bmy = \bPhi \bmx + \bmn =  \bTheta \bms +\bmn.
\end{align}
\revision{Here, the vector $\bmy\in\C^{M}$ contains all $M$ compressive measurements, the $i$th row of the sensing matrix $\bPhi^{M\times N}$ corresponds to the measurement vector $\bmphi_i$, the $M\times N$ effective sensing matrix $\bTheta=\bPhi\bPsi$ models the joint effect of CS and the sparsifying transform, and $\bmn\in\C^M$ models mesurement noise.}

The main goal of CS is to acquire far fewer measurements than the ambient dimension~$N$, i.e., we are interested in the case $M\ll N$; this implies that the matrix~$\bTheta$ maps $K$-sparse signals of dimension $N$ to a small number of measurements~$M$. Given a sufficient number of measurements, typically scaling as $M \sim K \log(N)$, that satisfy certain incoherence properties between the measurement matrix~$\bPhi$ and the sparsifying transform $\bPsi$, one can use sparse signal recovery algorithms that generate robust estimates of the sparse representation $\bms$ and hence, enable the recovery of the signal $\bmx=\bPsi\bms$   from the measurements in~$\bmy$; \revision{see~\cite{candes2008introduction,foucart2013mathematical} for more details on CS.}

\subsection{A2I Converter Architectures}
\label{sec:converterarchitectures}
\revision{While sparse signal recovery is typically carried out in software~\cite{maleki2010optimally} or with dedicated digital circuitry~\cite{maechler2012vlsi,bellasi2013vlsi}, the CS-based A2I conversion process modeled by \eqref{eq:acquisition} is implemented directly in the analog domain. The next paragraphs summarize the most prominent A2I converter architectures that perform CS measurement acquisition with mixed-signal circuitry.}
\subsubsection{Non-Uniform Sampling}
Non-uniform sampling (NUS) is one of the simplest instances of CS. In principle, the NUS strategy samples the incoming signal at irregularly spaced time intervals by taking a random subset of the samples of a conventional Nyquist ADC~\cite{wakin2012nonuniform,bellasi2013vlsi}. For this scheme, the sensing matrix~$\bPhi$ is given by the $M\times N$ restriction operator $\bR_\Omega=[\bI_N]_{\Omega,:}$ that contains of a subset $\Omega$ the rows of the identity matrix $\bI_N$, \revision{where $|\Omega|=M$ is the cardinality of the sampling set.}
The effective sensing matrix $\bTheta=\bR_\Omega\bI_N\bPsi$ in~\eqref{eq:acquisition} contains the $M$ rows of the sparsifying basis~$\bPsi$  indexed by~$\Omega$. 
More specifically, NUS can be modeled as 
\begin{align} \label{eq:NUS}
\bmy = \bR_\Omega \bI_N \bmx +\bmn = \bTheta_\text{NUS} \bms +\bmn
\end{align}
with $\bTheta_\text{NUS}= \bR_\Omega\bI_N \bF^H$, where we assume DFT sparsity.
\revision{As shown in \cite{foucart2013mathematical}, randomly-subsampled Fourier matrices enable faithful signal recovery from $M\sim K\log^4(N)$ compressive measurements. Hence, NUS not only enables sampling rates close to the Landau rate~\cite{landau1967sampling}, but is also conceptually simple.}

\begin{figure}[tp]
\centering
\includegraphics[width=0.8\columnwidth]{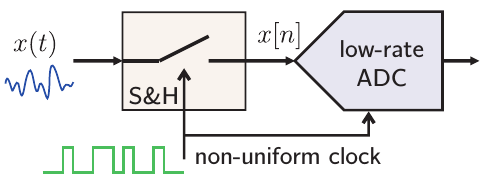}
\vspace{-0.1cm}
\caption{High-level architecture of non-uniform sampling (NUS). A sample-and-hold (S\&H) stage acquires a random subset of Nyquist-rate samples of a wideband signal $x(t)$ and converts each sample $x[n]$ to the digital domain.}
\label{fig:architecture_nus}
\end{figure}

A high-level architecture of NUS, as depicted in \figref{fig:architecture_nus}, consists of a sample-and-hold (S\&H) stage and an ADC supporting the shortest sampling period used by the NUS~\cite{wakin2012nonuniform,bellasi2013vlsi}. The main challenge of NUS is in the  
acquisition of a wideband analog input signal. While the average sampling rate can be  decreased significantly, the ADC still needs to acquire samples from wideband signals with frequencies potentially reaching up the maximal input signal frequency. 
This key observation has two consequences: First, NUS requires a sampling clock operating at the time resolution of the order of the Nyquist rate, which is typically power expensive. \revision{Second, NUS is sensitive to timing jitter: informally speaking, if the input signal changes rapidly, a small error in the sampling time can result in a large error in the acquired sample.}
\subsubsection{Variable Rate Sub-Nyquist Sampling}
Variable-rate sub-Nyquist sampling builds upon the fundamentals of bandpass sampling \cite{vaughan1991signal}. In principle, this A2I \revision{conversion} strategy undersamples the input signal with multiple branches \revision{(i.e., a bank of parallel bandpass sampling stages) with sampling rates that differ from one branch to the other.}
There exist two main instances of this concept, namely multi-rate sampling (MRS) that uses a fixed set of sampling frequencies for each branch~\cite{bai2012compressive,sun2013wideband,tzou2015low} and the Nyquist-folding receiver (NYFR) that modulates the sampling frequencies \cite{maleh2012analog}. 
Both approaches rely on the fact that the signal of interest is aliased at a particular frequency when undersampled at a given rate on a given branch, but the same signal may experience aliasing at a different frequency when sampled at a different rate on another branch. Empirical results show that this approach enables signal recovery for a sufficiently large number of branches~\cite{fleyer2010multirate}.

From a hardware perspective,  MRS is relatively simple as it avoids any randomness during the sampling stage and each branch performs conventional bandpass sampling. 
Nevertheless, MRS faces the same issues of traditional bandpass sampling~\cite{vaughan1991signal}: it suffers from noise folding, i.e., wideband noise in the signal of interest is folded (or aliased) into the compressive measurements, which results in reduced sensitivity~\cite{davenport2012pros,arias2011noise}.

%and \cs{why is that?} requires analog circuitry with stringent clock jitter constraints.
%
%
%\revision{The NYFR additionally requires means to accurately modulate the frequency of the sampling clock.}

%
\subsubsection{Random Modulation}
Random modulation (RM) is used by a broad range of A2I converters. Existing architectures first multiply the analog input signal by a pseudo-random sequence, integrate the product over a finite time window, and sample the integration result. 
The random-modulation preintegrator (RMPI) \cite{candes2008introduction,pareschi2015hardware} and its single branch counterpart, the random demodulator (RD) \cite{tropp2010beyond,chen2011sub}, are the most basic instances of this idea. 
\revision{However, modulating the signal with a \mbox{(pseudo-)}random sequence is only suitable for very specific signal classes, such as signals that are well-represented by a union of sub-spaces~\cite{mishali2011xampling}. In addition, the (pseudo-)random sequence generator must still run at Nyquist rate.}
\revision{The main advantage of the modulated wide-band converter (MWC) is to reduce the bandwidth of the S\&H to run at sub-Nyquist rates \cite{mishali2010theory,mishali2011xampling}.}
Indeed, the MWC avoids a fast sampling stage and, instead, requires a  high-speed mixing stage which is typically more wideband. 
A recent solution that avoids some of the drawbacks of RM is  the  quadrature A2I converter (QAIC)~\cite{yazicigil2015wideband}. \revision{This method relies on conventional down-conversion before RM, thus focusing on a small RF band rather than the entire  bandwidth.}

\subsection{Limitations of Existing A2I Converters}
\label{sec:limitations}
\revision{While numerous A2I converter architectures have been proposed in the literature, their limited practical success is a result of many factors. From a theoretical perspective, one is generally interested in acquisition schemes that minimize the number of  measurements while still enabling faithful recovery of a broad range of signal classes. From a hardware perspective, the key goals are to minimize the bandwidth requirements, the number of branches, and the power consumption, while being tunable  to the application at hand. Finally, suitable A2I converters should exhibit high sensitivity and be robust to hardware impairments and imperfections. We now summarize the key limitations of existing A2I converter architectures as discussed in \secref{sec:converterarchitectures} with these desirables in mind.}

Most of the discussed A2I converters rely on random mixing or sampling. Such architectures either require large memories to store the random sequences or necessitate  efficient means for generating pseudo-random sequences~\cite{rauhut2010compressive}.
In addition, such unstructured sampling schemes prevent the use of fast linear transforms (such as the fast Fourier transform) in the recovery algorithm, which results in excessively high signal processing complexity and power consumption~\cite{maechler2012vlsi,bellasi2013vlsi}. 
From a hardware perspective, large parts of the analog circuitry of many A2I converters must still support bandwidths up to the Nyquist rate, even if the average sampling rate is significantly reduced. For example, NUS \cite{wakin2012nonuniform} and MRS \cite{tzou2015low} require S\&H circuitry and ADCs designed for the full Nyquist bandwidth. 
Similarly, the RD and RMPI require sequence generators that run at the Nyquist rate.
Another drawback of many A2I converters, especially for MRS or the MWC \cite{mishali2010theory,mishali2011sub,mishali2011xampling,yazicigil201519}, is  that they require a large number of branches, which results in large silicon area and potentially high power consumption.

A more fundamental issue of most CS-based A2I converter solutions for wideband RF sensing applications is noise folding~\cite{davenport2012pros,arias2011noise}, which prevents their use  for applications requiring high sensitivity, such as activity detection of low-SNR signals.
In addition, most A2I converters lack versatility or adaptability to the application at hand, i.e., most system parameters are fixed at design time and signal acquisition is non-adaptive (one cannot select the next-best sample based on the history of acquired samples). However, adaptive CS schemes have the potential to significantly reduce the  acquisition time or the complexity of signal recovery \cite{malloy2014near}. 

% ================================================================================
% ================================================================================
% ================================================================================

\section{Non-Uniform Wavelet (Bandpass) Sampling}
\label{sec:NUWS}
We now propose a novel CS-based A2I converter that mitigates some of the drawbacks of existing A2I converter solutions. 
Our approach is referred to as non-uniform wavelet sampling (NUWS) and essentially acquires wavelet coefficients directly in the analog domain.
We first introduce the principle of NUWS and then adapt the method to multi-band  signals, resulting in non-uniform wavelet bandpass sampling (NUWBS). We then highlight the advantages of NUWS and NUWBS compared to existing A2I \revision{converters} for RF feature extraction.

\begin{figure}[t] % figures should generally be on top of the page; not within the text
\centering
\includegraphics[width=0.99\columnwidth]{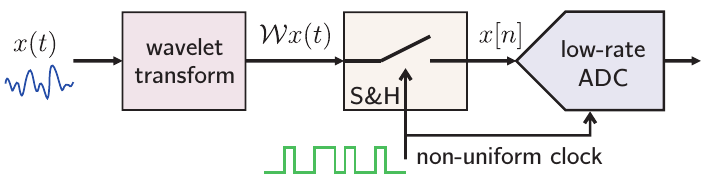}
\vspace{-0.4cm}
\caption{High-level architecture of non-uniform wavelet sampling (NUWS): %\cs{here describe the difference to NUS in one sentence:}%
Conceptually, NUWS first performs a continuous wavelet transform $\setW x(t)$ of the input signal $x(t)$, followed by NUS as shown in \figref{fig:architecture_nus} to obtain wavelet samples $x[n]$. A practical hardware architecture is discussed in \secref{sec:implementation}.}
\label{fig:architecture_nuws}
\end{figure}

\subsection{NUWS: Non-Uniform Wavelet Sampling}
\begin{figure}[t] % figures should generally be on top of the page; not within the text
\centering
\includegraphics[width=0.95\columnwidth]{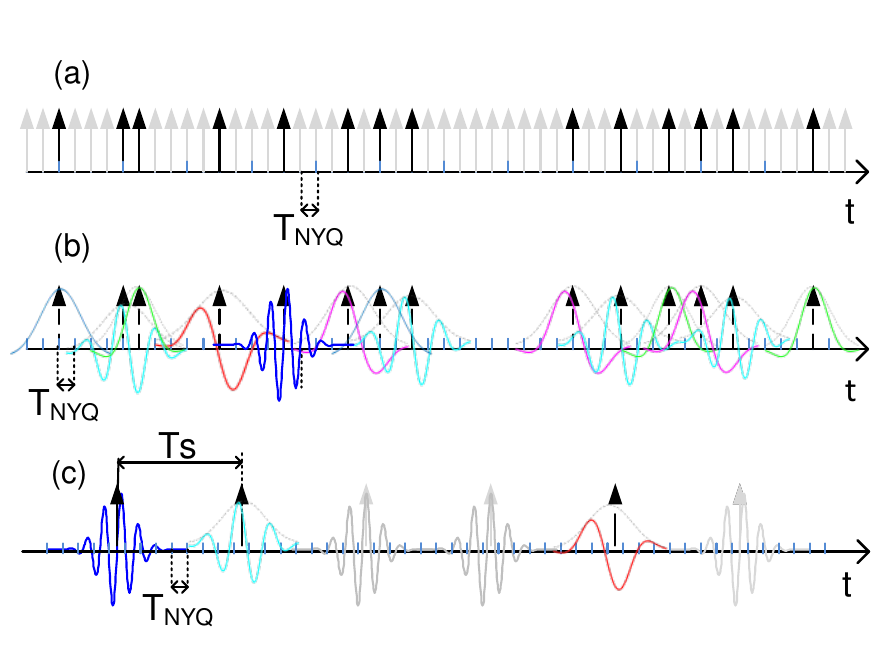}
\vspace{-0.2cm}
\caption{Illustration of the sampling patterns of NUS, NUWS, and NUWBS. NUS multiplies the incoming signals with a \revision{punctured} Dirac comb; NUWS multiplies the incoming signals with a series of carefully-selected wavelets; NUWBS uses a wavelet comb that is sub-sampled in time and with wavelets of variable central frequency in order to filter the sub-bands of interest.}
%In contrast to NUS shown in \figref{fig:architecture_nus}, NUWS projects the signal on custom wavelet before sub-sampling step.}
\label{fig:time_nus_nuws_nuwbs}
\end{figure}

The operating principle of NUWS is illustrated in \figref{fig:architecture_nuws}. In contrast to NUS (cf.~\figref{fig:architecture_nus}), NUWS first transforms the incoming analog signal $x(t)$ into a wavelet  frame $\setW x(t)$ (see \secref{sec:waveletbasics} \revision{for the basics on wavelets}) and then performs NUS to acquire a small set of so-called \emph{wavelet samples} $x[n]$. 
As illustrated in \figref{fig:time_nus_nuws_nuwbs}(a), NUS is equivalent to multiplying the input signal~$x(t)$ with a Dirac comb followed by the acquisition of a subset of samples (indicated by black arrows). \revision{In contrast, as shown in \figref{fig:time_nus_nuws_nuwbs}(b), NUWS multiplies the input signal~$x(t)$ with wavelets, integrates over the support of each wavelet, and samples the resulting wavelet coefficients.}

From a high-level perspective, NUWS has the following advantages over NUS. First, the continuous wavelet transform~$\setW$ reduces the bandwidth of the input signal~$x(t)$, which relaxes the bandwidth of the S\&H circuit and the ADC (see \secref{sec:NUWBSdesign} for the details).
Second, NUWS enables full control over a number of parameters, such as the sample time instants, wavelet bandwidth, and center frequency. In contrast, NUS has only one degree-of-freedom: the sample time instants.

In discrete time,  the sensing matrix $\bPhi$ for NUWS can be described by taking a small set $\Omega$ of rows of a (possibly overcomplete) wavelet frame $\bW^H\in\C^{W \times N}$, where~$\bW^H$ contains a specific wavelet on each row and \mbox{$W \geq  M$} corresponds to the total number of wavelets.  Hence, the sensing matrix of NUWS is  $\bPhi=\bR_\Omega\bW^H$, where $M=|\Omega|$ is the number of wavelet samples.
\revision{We can describe the NUWS process as}
\begin{align} \label{eq:NUWS}
\bmy = \bR_\Omega \bW^H \bmx + \bmn = \bTheta_\text{NUWS}\bms+\bmn
\end{align}
with the effective sensing matrix $ \bTheta_\text{NUWS}=\bR_\Omega\bW^H\bF^{H}$ where we, once again, assumed sparsity in the DFT domain.\footnote{Depending on the application, other sparsity bases $\bPsi$ than the \revision{inverse DFT matrix $\bF^{H}$} can be used; an investigation of other bases is ongoing work.} The necessary details on wavelets are provided in \secref{sec:waveletbasics}.

By comparing \eqref{eq:NUWS} with \eqref{eq:NUS}, we see that NUS subsamples the inverse DFT matrix, whereas NUWS subsamples the (possibly overcomplete) matrix~$(\bF\bW)^H$, which is the Hermitian of the Fourier transform of the entire wavelet frame. 
We can write the acquisition of the frequency-sparse signal $\bms$ as  
\begin{align} \label{eq:reformulatedNUWS}
\bmy = \bR_\Omega (\bF\bW)^H \bms + \bmn,
\end{align}
which implies that each wavelet sample is equivalent to an   inner product of the Fourier transform of the wavelet, i.e., $\hat\bmw_i=\bF\bmw_i$, with the sparse representation $\bms$ as $y_i=\langle\hat\bmw_i,\bms\rangle+n_i$, $i\in\Omega$.

\begin{figure}[t]  
\centering
\includegraphics[width=0.95\columnwidth]{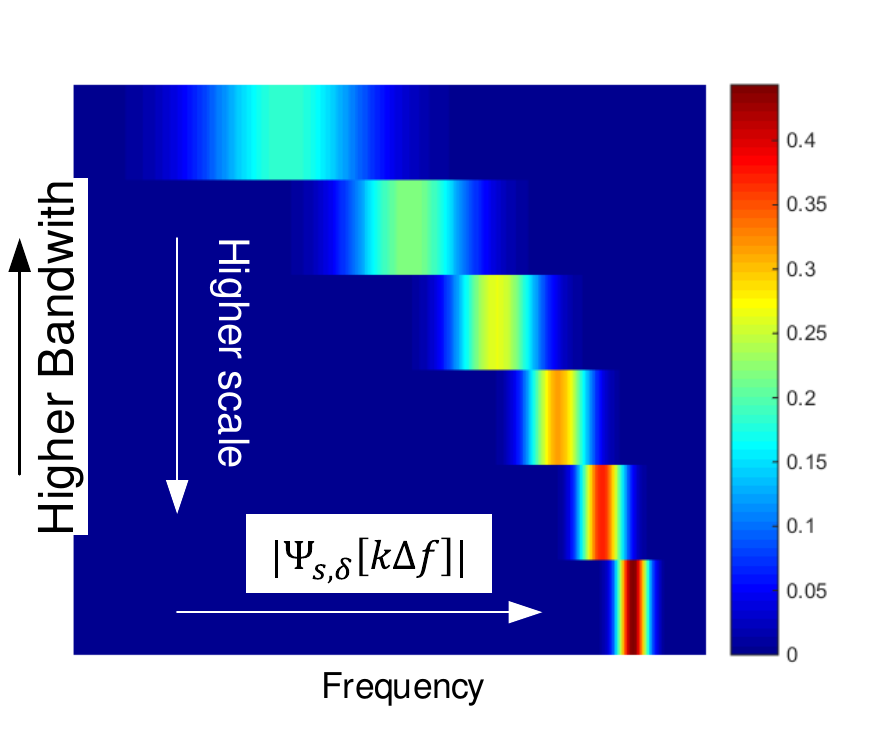}
\vspace{-0.4cm}
\caption{Absolute value of the product between the Hermitian of the complex-valued matrix and \revision{inverse DFT matrix  $|[\bW^H\bF^{H}]_{k,\ell}|$.} We see that each scale focuses on a different frequency band, whereas the bandwidth within each scale is fixed and the phase changes for different wavelets. }
\label{fig:wavelet_fourier_matrix}
\end{figure}

As we will discuss in detail in \secref{sec:waveletbasics}, the considered wavelets essentially correspond to bandpass signals with a given center frequency, bandwidth, and phase (given by the sample time instant). Thus, each wavelet sample corresponds to pointwise multiplication of the sparse signal spectrum with the bandpass filter equivalent to the Fourier transform of the wavelet. \figref{fig:wavelet_fourier_matrix} illustrates this property and shows the absolute value of the matrix $(\bF\bW)^H$ for the complex-valued Morlet wavelet~\cite{morlet1982wave} with \revision{six} scales.
Evidently, each wavelet captures a different portion of the spectrum with a different phase (phase differences are not visible) and bandwidth. We note that for the Morlet wavelet, the bandwidth and center frequency of each wavelet depends on the scale.

\subsection{NUWBS: Non-Uniform Wavelet Bandpass Sampling}
\label{sec:NUWBS}
\begin{figure}[tp]
\centering
\includegraphics[width=0.95\columnwidth]{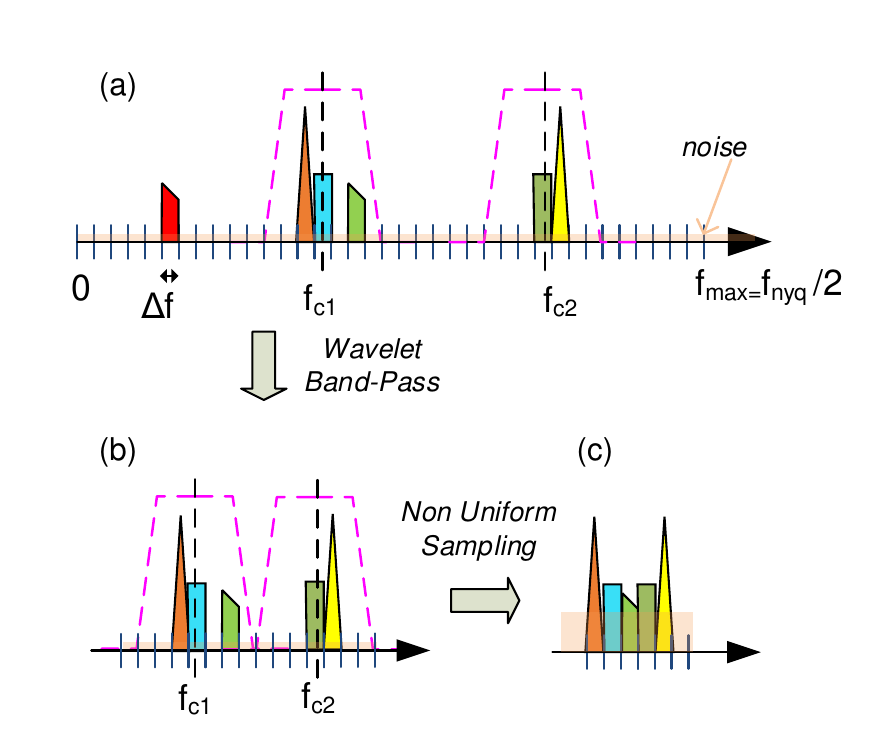}
\vspace{-0.4cm}
\caption{Illustration of a multi-band RF signal (a) consisting of two sparsely-populated sub-bands and an interferer (red). NUWBS first performs wavelet bandpass sampling to extract both sub-bands (b); then, NUS is used to minimize the number of wavelet samples, effectively reducing the sampling rate (c).}
\label{fig:spectrum_nuwbs}
\end{figure}

Non-uniform wavelet bandpass sampling (NUWBS) is a special instance of NUWS optimized for multi-band RF signals. 
The capability of handling such signals is of particular interest for non-contiguous carrier aggregation, a promising technology to enhance IoT throughput needs \cite{wunder20145gnow}.  
\figref{fig:spectrum_nuwbs}(a) illustrates a typical multi-band scenario in which RF signals occupy multiple non-contiguous frequency bands that may be sparsely populated; in addition, there may be interferers outside the sub-bands of interest.  
A standard way to acquire multi-band signals is to use a filterbank with one dedicated filter and RF receiver per sub-band. Besides requiring high complexity and power, and suffering from lack of flexibility, such designs are typically unable to exploit signal sparsity within the sub-bands.

Traditional bandpass sampling  \cite{vaughan1991signal} or NUS \cite{venkataramani2000perfect,venkataramani2001optimal,lazar2004perfect,mishali2009blind} for multi-band signals would result in several issues. 
\revision{First and foremost, noise and interferers outside the sub-bands of interest will inevitably fold (or alias) into the measurements---a phenomenon known as noise folding \cite{vaughan1991signal,arias2011noise,davenport2012pros}.}
Furthermore, for NUS, the a-priori information on the occupied sub-bands is generally not exploited during the acquisition process.

In stark contrast to these methods, NUWBS exploits the multi-band structure and sparsity within each sub-band, while being resilient to interferers or noise outside of the bands of interest. 
As illustrated in \figref{fig:time_nus_nuws_nuwbs}(c), NUWBS multiplies the incoming signals with a wavelet comb on a regular sampling grid, sub-sampled in time with respect to the Nyquist rate.
Unlike NUWS, there are no overlaps between wavelets, which prevents the need for a large number of branches---typically one branch per sub-band is sufficient. 
\revision{Furthermore, the center frequencies of the wavelets can be focused on the sub-bands of interest, which renders this approach resilient to out-of-band noise and interferers, effectively reducing noise folding and aliasing without the need for a filter bank.}
Finally, as shown in \secref{sec:numericalresults}, NUWBS is able to leverage CS and achieves near-optimal sampling rates, i.e.,  close to the Landau rate.

\revision{The operating principle of NUWBS is illustrated in \figref{fig:spectrum_nuwbs}. Every NUWBS measurement acts like a filter, which removes out-of-band noise and interference (see  \figref{fig:spectrum_nuwbs}(b)). Then, as shown in \figref{fig:spectrum_nuwbs}(c), by taking a subset of wavelet samples (i.e., wavelet bandpass sampling), NUWBS reduces the average sampling rate. Traditional recovery methods for CS can then be used to recover the multi-band signals of interest.
From a mathematical viewpoint, NUWBS can be modeled as in~\eqref{eq:NUWS} with the differences that the subset of samples $\Omega$ is adapted to the sub-bands of interest and the wavelet samples are on a regular sub-sampled grid with non-overlapping wavelets.} 

\subsection{Advantages of NUWS and NUWBS}
Wavelets find broad applicability in wireless communication systems, including source coding, modulation,  interference mitigation, and signal de-noising~\cite{lakshmanan2006review,nikookar2013wavelet}.
Nevertheless, CS-based methods that rely on wavelet sampling are rather unexplored, especially when dealing with RF signals. 
A notable exception is the paper~\cite{matusiak2012sub}, in which a multi-channel acquisition scheme based on Gabor frames is proposed that exploits the sparsity in the time-frequency domain. 
In contrast to NUBS/NUWBS, this approach relies on a parallel set of Gabor sampling branches, where each Gabor wavelet has a fixed bandwidth and the sampling rate is reduced using the MWC.

\revision{We next summarize the benefits of wavelet sampling and  the advantages of NUWS/NUWBS to RF applications.}

\subsubsection{Tunability and Robust Feature Acquisition}
Wavelets offer a broad range of parameters including time instant,  center frequency, and bandwidth (see \secref{sec:waveletbasics} for the details). This flexibility can be exploited to adapt each measurement to the signal or feature class at hand or to improve robustness to out-of-band noise and interferers,   or aliasing. \revision{For NUWBS, we take advantage of this property by focusing each wavelet sample on the occupied sub-bands, which yields improved sensitivity by mitigating noise folding and interference.}

\subsubsection{Adaptive Feature Extraction} 
\revision{The tree structure of wavelets across scales~\cite{ariananda2009study} is a well-exploited property in data compression~\cite{boliek2001jpeg2000}. In RF applications, one can exploit this property to develop adaptive feature extraction schemes that first identify RF activity on a coarse scale (e.g., in a wide frequency band) and then, adaptively ``zoom in'' to sub-bands that exhibit activity for a more detailed analysis. This approach avoids traditional frequency scanning  and  has the potential to enable faster RF feature extraction than non-adaptive schemes.}

\subsubsection{Structured Sampling} 
A broad range of CS-based A2I \revision{converter solutions focuses} on randomized or unstructured sampling methods. Such methods typically require large storage (for the sampling matrices) and high complexity during signal recovery. In contrast, structured sensing approaches are known to avoid these drawbacks~\cite{duarte2011structured}. Wavelets exhibit a high degree of structure and their parametrization requires low storage. \revision{Furthermore, recovery algorithms that rely on fast (inverse) wavelet transforms are computationally efficient~\cite{soma2008characterization}.}

\subsubsection{Relaxed Hardware Constraints}
From a hardware perspective, random sequences or clock generation circuitry that operates at Nyquist rates can---in contrast to NUS and RD---be avoided  \revision{due to the sub-Nyquist operation of} NUWS and NUWBS. \revision{Hence, the associated clock synthesis and clock-tree management can be relaxed}  \cite{devries2008subsampling}.
In addition, by sub-sampling the wavelet coefficients, we can further reduce the ADC sampling rates. \revision{Due to the signal correlation with the wavelet prior to sampling}, the bandwidth requirements of the S\&H circuit and the ADC \revision{are relaxed as well}.
In addition, NUWBS prevents overlapping wavelets, which enables the use of a small number of parallel sampling branches. 
This property reduces the circuit area and power consumption. As we will show in \secref{sec:implementation}, \revision{widely-tunable wavelets can be generated efficiently in analog hardware.}

% ================================================================================
% ================================================================================
% ================================================================================

\section{Wavelet Design and Validation of NUWBS}
\label{sec:NUWBSdesign}
\revision{This section summarizes the basics of wavelets and then, discusses wavelet selection and design for NUWS/NUWBS. We finally validate NUWBS for multi-band RF sensing.}

\subsection{Wavelet Prerequisites}
\label{sec:waveletbasics}
For the sake of simplicity, we will  use both continuous-time and discrete-time signal representations and often switch in between without making the discretization step explicit.
 
\subsubsection{Wavelet Basics}
A wavelet is a continuous waveform that is effectively limited in time,  has an average value of zero, and bounded $L^2$-norm (often normalized to one). Wavelets for 
signal processing were introduced by Morlet \cite{morlet1982wave} who showed that continuous-time functions $x(t)$ in $L^2$ can be represented by a so-called wavelet $\psi_{s,\delta}(t)$ that is obtained by scaling~$s\in\R^+$ and shifting~$\delta\in\R$ a so-called mother wavelet $\psi(t)$. The scaling and shifting   operations can be \revision{made formal} as follows:
\begin{align} \label{eq:waveletdefinition}
\psi_{s,\delta}(t)= \frac{1}{\sqrt{s}}\psi\!\left(\frac{t-\delta}{s}\right)\!, \quad s\in\R^+,\,\delta\in\R.
\end{align}
The so-called wavelet coefficient $\setW x_{s,\delta}$ of a signal~$x(t)$ for a given wavelet $\psi_{s,\delta}(t)$ at scale $s$ and with time shift $\delta$, is defined as the following inner product~\cite{mallat1999wavelet,daubechies1992ten}:
\begin{align} \label{eq:waveletcoefficient}
\setW x_{s,\delta} = \langle x , \psi_{s,\delta}\rangle = \int_{\R} x(t) \frac{1}{\sqrt{s}}\psi^*\!\left(\frac{t-\delta}{s}\right)\! \text{d}t.
\end{align}
In words, each wavelet coefficient $\setW x_{s,\delta}$ compares the signal~$x(t)$ to a shifted and scaled version~$\psi_{s,\delta}(t)$ of the mother wavelet $\psi(t)$. 
By comparing the signal to wavelets for various scales and time shifts, we arrive at the \emph{continuous wavelet transform (CWT)} $\setW x_{s,\delta}$. The CWT represents one-dimensional signals in a highly-redundant manner, i.e., by two continuous parameters $(s,\delta)$. All possible scale-time atoms can be collected in an  (overcomplete) frame given by
\begin{align*} %\label{eq:waveletdict}
\setD = \left\{\psi_{s,\delta}(t) \,\big|\,  \delta\in\R, s\in\R^+\right\}\!.
\end{align*}
In practice, one is often interested in selecting a suitable subset of scales and shifts that enable an accurate (or exact) representation of  original signal $s(t)$ of interest. In what follows, we are particularly interested in wavelets that can be generated efficiently in hardware; such wavelets are discussed next.

\subsubsection{Gabor Frame}
\begin{figure}[t] % figures should generally be on top of the page; not within the text
\centering
\includegraphics[width=0.95\columnwidth]{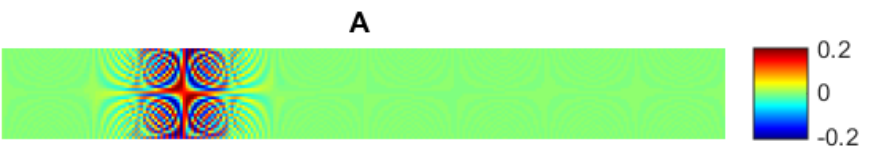}
\caption{\revision{Real part of the product between the Hermitian of the Gabor and inverse DFT matrix $\bA=\Re\{\bW^H\bF^{H}\}$ at a given center frequency and for various time shifts. Signals far away from the center frequency are attenuated, which effectively mitigates out-of-band noise, interference, and aliasing.}}
\label{fig:frequency_rejection}
\end{figure}

The Gabor transform is a well-known analysis tool to represent a signal simultaneously in time and frequency, similarly to the short-time Fourier transform (STFT). The set of Gabor functions (often called Gabor frame) is, strictly speaking, not a wavelet basis---the formalism, however, is very similar \cite{benedetto1998gabor,christensen2006pairs}.
Gabor frames consist of functions (or atoms)
\begin{align} \label{eq:gabor}
\psi_{f^c_{\nu},\delta_k}(t) = p(t-\delta_k)e^{j2\pi f^c_{\nu} t},
\end{align} 
which are parametrized by the center frequencies $f^c_{\nu}$ and time shifts $\delta_k$ of a windowing function $p(t)$, where $\nu=1,2,\ldots$ and $k=1,2,\ldots$ are the indices of discrete frequency and time shifts, respectively.
In practice, one often uses a Gaussian windowing function $p(t)$ that is characterized by the \revision{width (or duration) parameter~$\tau$.}
Based on \cite{benedetto1998gabor}, the time and frequency representation of the unit $\ell_2$-norm Gabor atoms with a Gaussian window are defined as follows:  
\begin{align} 
\psi_{f^c_{\nu},\delta_k}(t) & = \frac{2^{\frac{1}{4}}}{\sqrt{\tau}\pi^{\frac{1}{4}}} e^{j2\pi f^c_{\nu}(t-\delta_k)}e^{-\left(\frac{t-\delta_k}{\tau}\right)^2}\label{eq:wavelettime} \\
\Psi_{f^c_{\nu},\delta_k}(f) & = (\tau\sqrt{2\pi})^\frac{1}{2} e^{-j2\pi\delta_k f} e^{-(\pi\tau(f-f^c_{\nu}))^2}. \label{eq:waveletfrequency}
\end{align}
There exists a trade-off when choosing the width parameter~$\tau$: a large width increases the frequency resolution while lowering the time resolution, and vice versa. 
\revision{As it can be seen from \eqref{eq:waveletfrequency}, the Fourier representations of Gabor atoms decay exponentially fast, which is the reason for their excellent frequency-rejection properties, i.e., signals sufficiently far away of the center frequency $f^c_{\nu}$ are strongly attenuated.
This filtering effect of Gabor atoms is illustrated in \figref{fig:frequency_rejection}, which shows the real part of the matrix~$(\bF\bW)^H$ for one particular center frequency $f^c_{\nu}$ and various time shifts $\delta_k$. Clearly, signals that are sufficiently far apart from the center frequency $f^c_{\nu}$ will be  filtered.}

\subsubsection{Complex-Valued Morlet Wavelet (C-Morlet)}
\begin{figure}[t] % figures should generally be on top of the page; not within the text
\centering
\includegraphics[width=0.95\columnwidth]{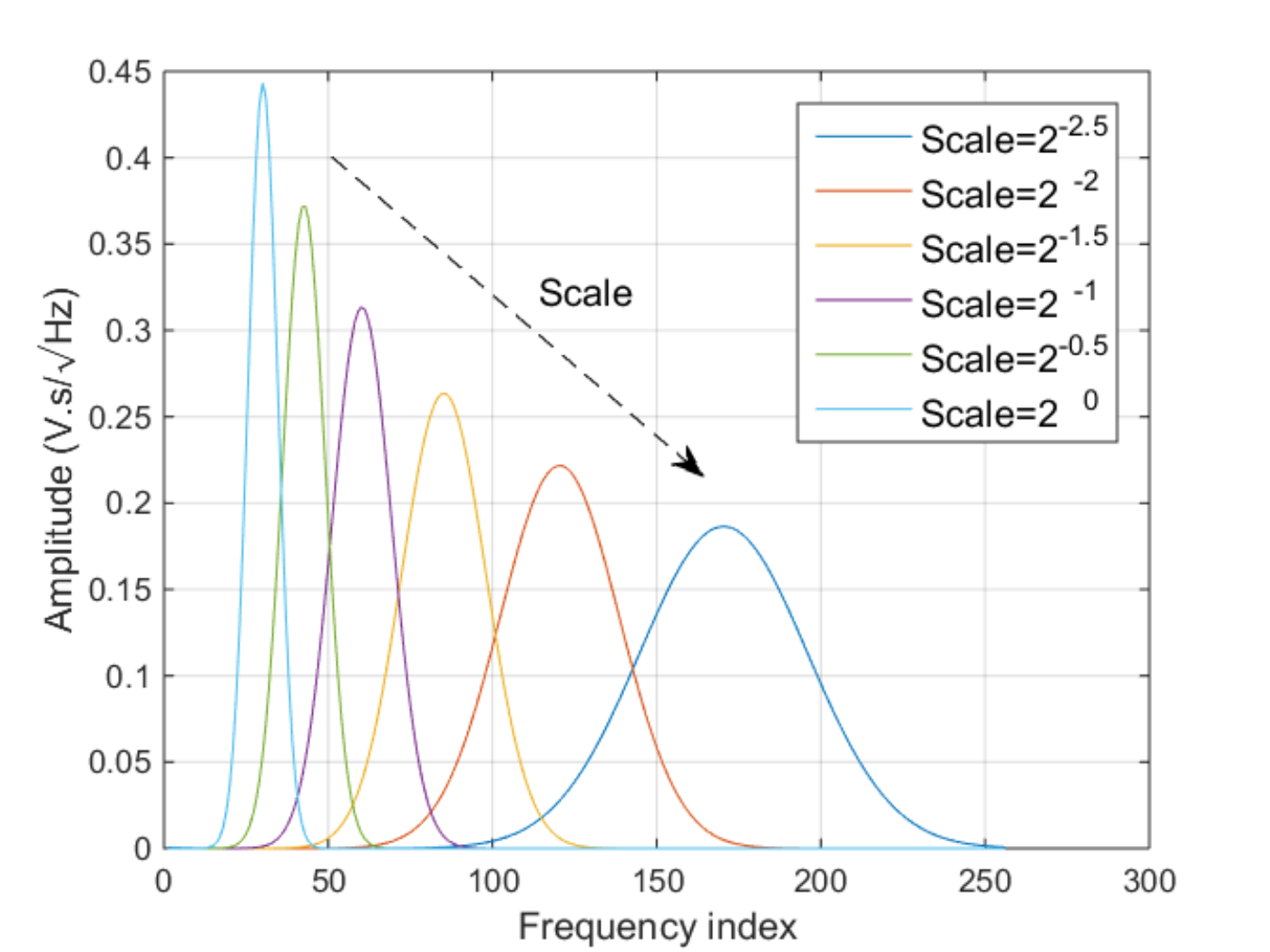}
\caption{Frequency domain amplitude of C-Morlet wavelets for $6$ scales as shown in \figref{fig:wavelet_fourier_matrix}; the bandwidth of the wavelets increases with the central frequency, which is in contrast to Gabor atoms that have constant bandwidth.}
\label{fig:cmorlet}
\end{figure}

In contrast to the Gabor frame, the complex-valued Morlet (C-Morlet) wavelet uses windowing functions whose width parameter is linked to the central frequency  (cf.~\figref{fig:wavelet_fourier_matrix})~\cite{mallat1999wavelet,daubechies1992ten}. 
Recall from \eqref{eq:waveletdefinition} that higher scales correspond to the most ``stretched'' wavelets (in time) and hence, wavelets measure long time intervals for features containing low-frequency information and shorter intervals for high-frequency information.
In fact, the width of a C-Morlet is linked to the central frequency so that there is a constant number of oscillations per effective wavelet duration. 
More formally, the C-Morlet shows a constant quality factor~$Q$ across scales. 
\revision{As a result, the C-Morlet wavelets coincide with~\eqref{eq:wavelettime} and satisfy the additional constraint accross scales that the central frequency $f^c_{\nu}$ and the wavelet bandwidth $\textit{BW}_p$ satisfy the following condition:}
%\begin{align*}
$Q = {f^c_{\nu}}/{\textit{BW}_p} = f^c_{\nu} \tau_\nu \pi \alpha \sqrt{2}$.
%\end{align*}
%
Here, the parameter $\alpha$ is $0.33$ for a $-10$\,dB referred bandwidth. \figref{fig:cmorlet} shows the spectrum amplitude for six scales of the C-Morlet wavelet family for a given quality factor---clearly the wavelet bandwidth is linked to the scale.

\subsection{Wavelet Selection for the Design of NUWBS}
\label{sec:atomselection}
\revision{We are, in principle, free in choosing the width, frequency, and time instant of each wavelet. In practice, however, we are interested in wavelets that can be generated efficiently in hardware, enable the use of a small number of branches, and extract the RF features of interest.} We now outline how to select suitable wavelet parameters for NUWBS.

\subsubsection{\revision{Parameter Selection}}
As detailed in \secref{sec:NUWBS}, NUWBS first performs a projection of the input signal on a select set of wavelets (or atoms) and then, subsamples the wavelet coefficients. 
Since Gabor frames contain a highly redundant set of atoms, \revision{it may---at first---seem counter-intuitive} to use an overcomplete frame expansion $\bW^H\in\C^{W\times N}$ with $W\gg N$ of the signal $\bmx$ as our ultimate goal is to reduce the number of measurements. 
It is thus critical to select a suitable subset~$\Omega$ of atoms that enables robust signal recovery or feature extraction with a minimum number of measurements $M=|\Omega| \ll N \ll W$. As we will see, the high redundancy turns out to be beneficial as it allows us to select a potentially better subset of measurement, e.g., compared to NUS that can only select a subset of rows of the Fourier matrix. 

If we are interested in optimizing our set $\Omega$ of wavelet samples for sparse signal recovery, which is the original motivation of CS, then we can minimize the so-called \emph{mutual coherence}~\cite{candes2008introduction} between the sub-sampled sensing matrix  $\bR_\Omega\bW^H$ and the sparsifying basis $\bPsi=\bF^{H}$, defined as
\begin{align} \label{eq:fullmutualcoherence}
\mu_m(\bR_\Omega\bW^H,\bF^{H}) = \max_{i,k} |\langle [\bR_\Omega\bW]_i,[\bF^{H}]_k\rangle|.
\end{align}
The mutual coherence is related to the minimum number of measurements $M$ that are required to guarantee recovery of $K$-sparse signals~\cite{candes2007sparsity,candes2008introduction}. \revision{Hence, we wish to find an optimal set~$\Omega$ of cardinality $M$ that minimizes \eqref{eq:fullmutualcoherence}; unfortunately, this is a combinatorial optimization problem. We therefore resort to a qualitative analysis and heuristics to identify a suitable set of wavelets that enables the recovery of sparse signals.}

According to the closed-form expression in \eqref{eq:wavelettime}, the Gabor atoms are characterized by the width parameter $\tau$. Our goal is to find the optimal width parameter $\hat\tau$, depending on the input signal (e.g., its bandwidth).
Intuitively, the width parameter~$\tau$ should be linked to the bandwidth $\textit{BW}_\text{RF}$ of the RF signal.
In fact, the effective wavelet width should be designed so that each wavelet measurement $y_i$, $i=1,2,\ldots,M$, collects enough information over the bandwidth of interest or, in other words, the pulse spectrum should be as flat as possible over the bandwidth of interest.
\revision{We can make this intuition more formal by considering the so-called \emph{local} mutual coherence~\cite{candes2006robust,krahmer2014stable} 
\begin{align} \label{eq:localmutualcoherence}
\mu_m(\bW^H_s,\bF^{H}_\Sigma) = \max_{i,k} |\langle [\bW_s]_i,[\bF^{H}_\Sigma]_k\rangle|
\end{align}
between the wavelet sampling matrix $\bW_s^H$  at a particular scale~$s$ and the sparsifying basis limited to the subset of frequencies~$\Sigma$ of interest (e.g., limited to the potentially active or occupied sub-bands).}

\begin{figure}[t] % figures should generally be on top of the page; not within the text
\centering
\includegraphics[width=0.95\columnwidth]{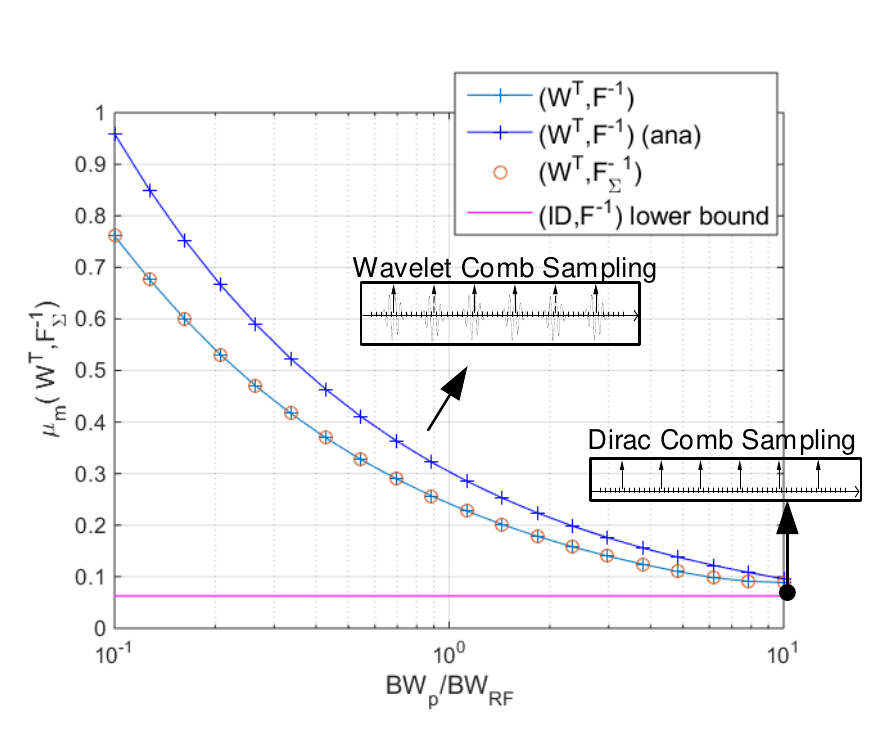}
\vspace{-0.5cm}
\caption{\revision{Mutual coherence $\mu_m$ between the Gabor frame $\bW^H$ and the inverse DFT matrix $\bF^{H}$ as a function of the wavelet bandwidth ($\textit{BW}_p$) relative to the RF signal bandwidth $\textit{BW}_p/\textit{BW}_\text{RF}$.}}
\label{fig:mutualcoherence}
\end{figure}

From the Gabor frame definition in  \eqref{eq:wavelettime}, we can compute a closed form expression of the mutual coherence between a Gabor frame having a fixed width parameter $\tau$ and the Fourier basis. Assuming that the atom's central frequency $f^c_{\nu}$ is centered to the band of interest and is a multiple value of the frequency resolution $\Delta f=f_\text{Nyq}/N$, we can compute the local mutual coherence defined in \eqref{eq:fullmutualcoherence} as follows:  
\begin{align*}
\mu_m(\bW^H_s,\bF^{H}_\Sigma) =  (\tau\sqrt{2\pi})^{1/2} 
\end{align*}
\figref{fig:mutualcoherence}  shows the evolution of the mutual coherence as well as the theoretical lower bound (the purple horizontal line) given by~$1/\sqrt{N}$~\cite{tropp2007signal,rauhut2010compressive}. \revision{The curves in this figure are obtained by  setting the dimension to $N=256$ and computing inner products between the rows of Gabor frame and the rows of the inverse discrete Fourier restricted to the band of interest~$\bF^{H}_\Sigma$.
The (local) mutual coherence is then computed according to \eqref{eq:fullmutualcoherence} (and Eq.~\ref{eq:localmutualcoherence}). The individual points on the curves are obtained by tuning the wavelet bandwidth $\textit{BW}_p$ divided by the occupied RF bandwidth~$\textit{BW}_\text{RF}$.} 
Note that the shorter the atom (or wavelet) duration $\tau$, the wider its bandwidth $\textit{BW}_p$ is. 
%\cs{I cannot really follow your line of arguments here: can you clarify?} 
As a result, the energy of the sensing vector is spread in the frequency domain and hence, captures  information of all frequencies within the sub-band of interest. The limit $\tau\rightarrow0$ corresponds to the Dirac comb \revision{(the bandwidth tends to infinity)} for which the mutual coherence is known to reach the Welch lower bound~\cite{studer2012recovery}.  
\revision{The limit $\tau \to \infty$ corresponds to the case in which the sensing vectors are localized in the frequency domain, i.e., the measurements are maximally coherent with the Fourier basis.} Hence, for wavelet sampling, we can determine the width parameter $\tau$ to match the signal of interest. In practice, one can trade-off filtering performance (to mitigate noise folding and aliasing) versus measurement incoherence (to reduce the number of required CS measurements).

\subsubsection{Gabor Time-Shift Selection}
\begin{figure}[t] % figures should generally be on top of the page; not within the text
\centering
\includegraphics[width=0.95\columnwidth]{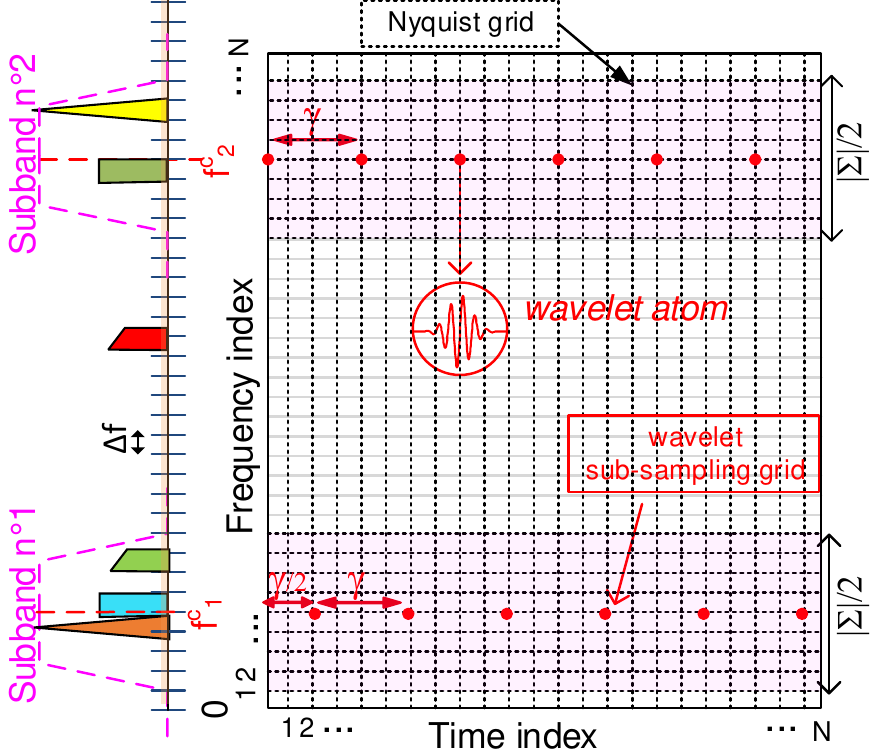}
\caption{\revision{Time-frequency grid of the Gabor atoms to be acquired via NUWBS; our approach makes use of a-priori knowledge of the occupied frequency bands; Atom selection relies on constant frequency and bandpass sampling in time for 2 sub-bands; the used parameters are $\textit{BW}_\text{ag}=2\times\textit{BW}_\text{RF} = 2\times16\Delta f$, $\gamma=16$, and $N=256$ samples.}}
\label{fig:selection}
\end{figure}

Besides selecting the optimal width parameter $\tau$ of the Gabor atoms, we have to identify suitable frequencies $f^c_\nu$ and time shifts~$\delta_k$. 
Consider, for example, the multi-band signal shown \revision{on the left side in~\figref{fig:selection}, where we assume that we know the coarse locations of the potentially active sub-bands (e.g., determined by a given standard), but not the locations of the non-zero frequencies within each sub-band (e.g., the frequency slots used for transmission). For simplicity, the axes have been normalized so that the y-axis stands for the frequency index $\nu$ and the x-axis stands for the time shift index~$k$.
We define the following parameters:  the sub-sampling ratio $\gamma=f_\text{Nyq}/\textit{BW}_\text{RF}$ is the ratio between the Nyquist frequency $f_\text{Nyq}$ and the bandwidth of each sub-band $\textit{BW}_\text{RF}$; the aggregate bandwidth $\textit{BW}_\text{ag}$ is the total bandwidth of all occupied sub-bands, i.e., in our example $\textit{BW}_\text{ag}=2  \textit{BW}_\text{RF}$. Equivalently, the aggregate bandwidth can be expressed by the cardinality of the occupied frequency indices~$|\Sigma|$ so that  $\textit{BW}_\text{ag}=\Delta f |\Sigma|$, where $\Delta f$ is the bandwidth per frequency bin.}

Our proposed Gabor frequency and time shift selection strategy relies on two principles. 
First, in order to acquire information in a given sub-band, we consider a fixed central frequency centered in the sub-band of interest. 
Second, in the time domain we perform bandpass sampling with the goal of mixing the signal of interest to (or near to) baseband.
This means that instead of sampling at all of the available time shifts defined by the Nyquist rate (shown by the vertical black dashed lines in \figref{fig:selection}), we only acquire a subset defined by the sub-sampling factor $\gamma$ (the red circles in \figref{fig:selection}), effectively performing wavelet bandpass sampling.
As a result two adjacent Gabor atoms will not overlap in time since, by construction, the sampling rate is inversely proportional to the pulse duration.  
In the case of two sub-bands the aggregate sampling rate is set to $2 f_\text{Nyq}/\gamma$ equivalent to the aggregate bandwidth equal to $2\textit{BW}_\text{RF}$. 
\revision{We note that in addition to bandpass sampling, we can perform NUS on the acquired wavelet samples to further reduce the sampling rates. As shown next, this is typically feasible in the case where we know a-priori that the sub-bands are sparsely populated.}

\subsection{Performance Validation of NUWBS}
\label{sec:numericalresults}
\revision{We now demonstrate the efficacy of NUWBS for spectral activity detection in a multi-band RF application.
In particular, we simulate an empirical phase transition~\cite{donoho2009message,maleki2010optimally} that characterizes the probability of correct support recovery, i.e., the rate of correctly recovering the true active frequency bins from NUWBS measurements. 
As a reference, we also include the theoretical phase transition of $\ell_1$-norm based sparse signal recovery for a Gaussian measurement ensemble~\cite{donoho2009message}.}

\begin{figure}[t] % figures should generally be on top of the page; not within the text
\centering
\includegraphics[width=0.95\columnwidth]{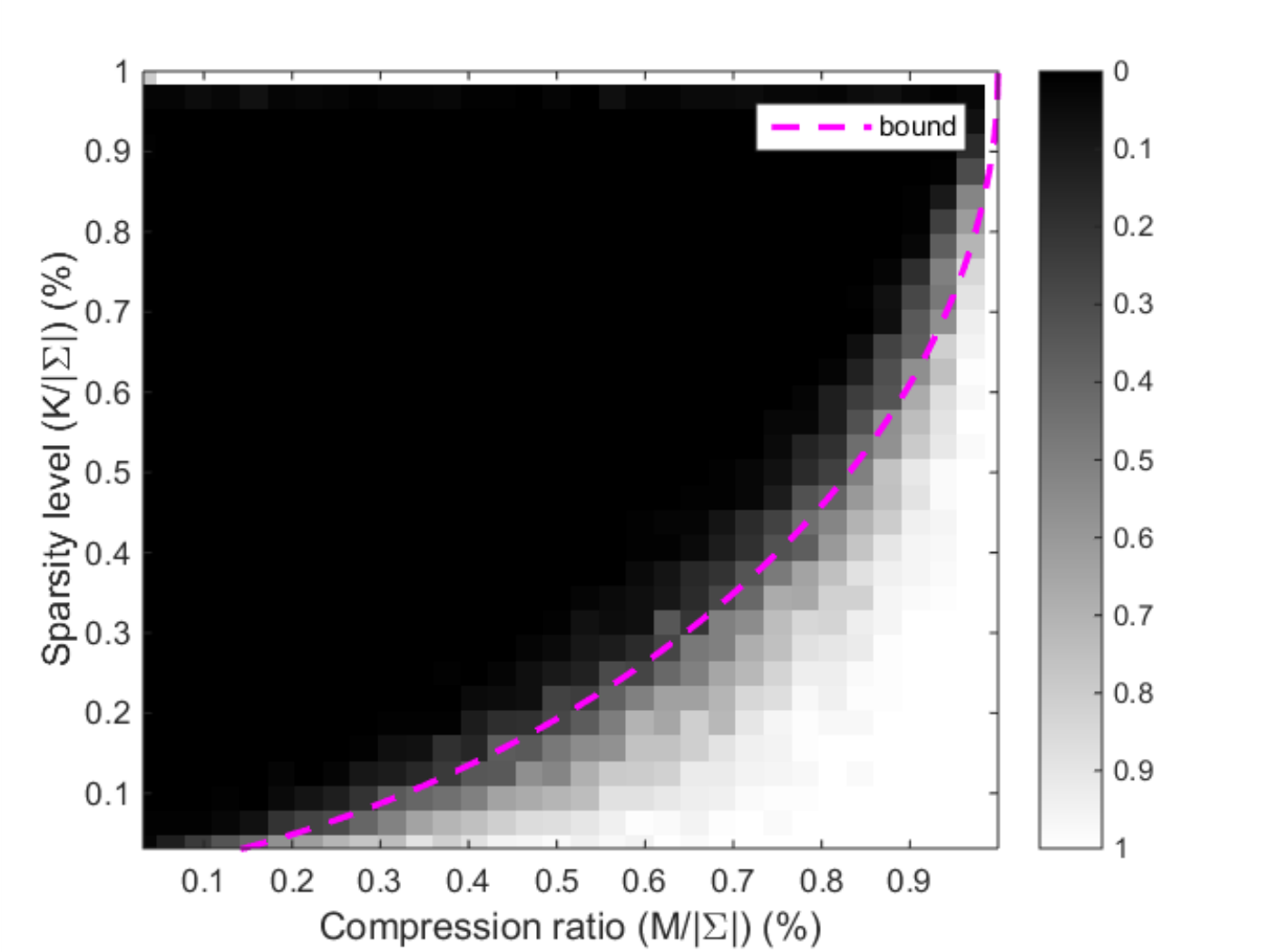}
\caption{Empirical phase transition graph of NUWBS for multi-band signal acquisition compared to the theoretical $\ell_1$-norm phase transition for a Gaussian measurement ensemble (shown with the dashed \revision{purple} line). NUWBS exhibits similar performance as the theoretical phase transition, which demonstrates that NUWBS enables near-optimal sample rates.}
\label{fig:phasetransition}
\end{figure}

\revision{We use $N=256$ frequency bins and two active sub-bands with a total number of $|\Sigma|=32$ potentially active frequency bins. The signals within these bins are assumed to be $K\leq |\Sigma|$ sparse.  
The NUWBS measurements are selected as discussed in \secref{sec:atomselection} and illustrated in \figref{fig:selection}, i.e., we form the $M\times N$ matrix $\bTheta_\text{NUWBS}=\bR_\Omega \bW^H\bF^{H}$ by fixing the frequency~$f^c_\nu$ at the center of each sub-band and use a sub-sampling ratio per branch of $\gamma=2N/|\Sigma|=16$.
We generate measurement-sparsity pairs $(M,K)$, and for each pair, we generate a synthetic $K$-sparse signal within the two allowed sub-bands; the $K$ non-zero coefficients are complex-valued numbers of unit amplitude and random phases.  
For support recovery, we use orthogonal matching pursuit~\cite{tropp2007signal}, restricted to the sub-bands of interest, i.e., we assume that the sub-band support $\Sigma$ is known a-priori but not the active coefficients within these sub-bands. We perform support set recovery for $100$ Monte--Carlo trials and report the average success rate.}

\revision{\figref{fig:phasetransition} shows the empirical phase transition, where white areas indicate zero errors for support set recovery. The x-axis shows the normalized compression ratio, i.e., the number of measurements compared to the total sub-band width $M/|\Sigma|$; the y-axis shows the normalized sparsity level, i.e., the fraction of non-zeros compared to the total sub-band width $K/|\Sigma|$.
We see that NUWBS exhibits a similar success-rate profile as predicted by the theoretical phase transition (i.e., recovery will fail above and succeed below the dashed purple line), which is valid in the asymptotic limit for $\ell_1$-norm based sparse-signal recovery from Gaussian measurements. 
This key observation implies that NUWBS in combination with the atom selection strategy discussed in 
\secref{sec:atomselection} exhibits near-optimal sample complexity  in multi-band scenarios.
We emphasize that even for the relatively small dimensionality of the simulated system (i.e., $N=256$), NUWBS is already in satisfactory agreement with the theoretical performance limits for sparse signal recovery.}

% ================================================================================
% ================================================================================
% ================================================================================

\section{Implementation Aspects of Non-Uniform Wavelet (Bandpass) Sampling}
\label{sec:implementation}
\revision{This section discusses hardware implementation aspects to highlight the practical feasibility of NUWS/NUWBS and their advantages over existing A2I converter solutions.}

\subsection{Architecture Considerations of NUWBS}
\begin{figure}[t] % figures should generally be on top of the page; not within the text
\centering
\includegraphics[width=0.95\columnwidth]{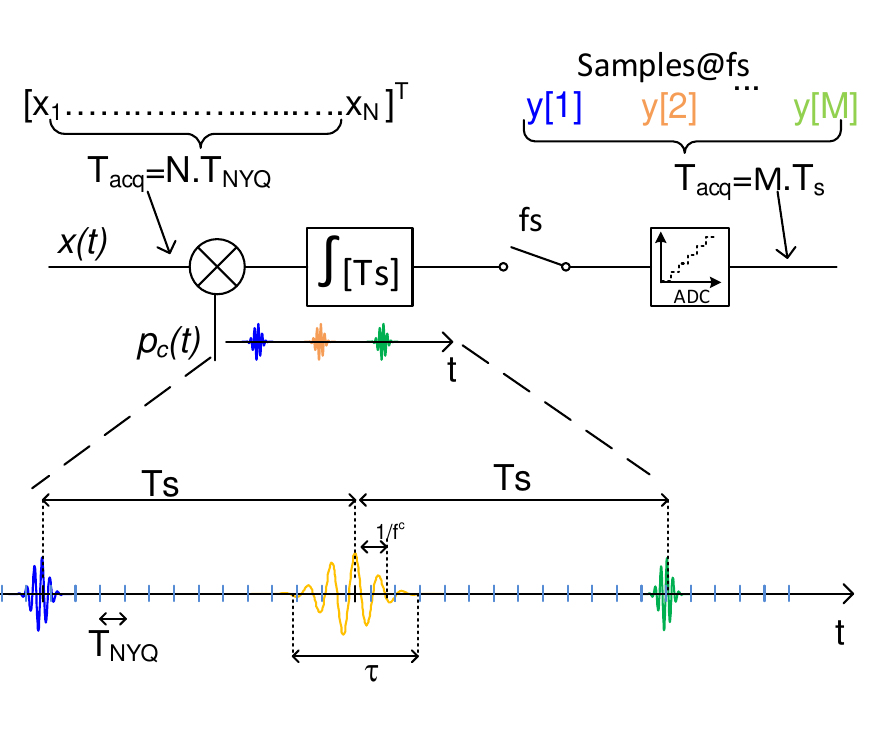}
\vspace{-0.7cm}
\caption{\revision{Generic serial NUWBS architecture to acquire Gabor frame or wavelet samples. NUWBS first multiplies the input signal $x(t)$ with a wavelet comb $p_c(t)$ at rate $1/T_s$ and integrates the result. One then takes a random set of wavelet samples and quantizes them using an ADC. A wavelet is defined by its central frequency $f^c$ and width parameter $\tau$ (the effective pulse duration).}}
\label{fig:implementation}
\end{figure}
\figref{fig:implementation} shows the critical architecture details for NUWBS that uses Gabor frames or C-Morlet wavelets. 
The continuous-time input signal~$x(t)$ is first multiplied (or mixed) with a wavelet comb  $p_c(t)$.
The resulting signal is then integrated over a period $T_s$ (for each wavelet) and subsampled at a rate~$f_s$.
The rate $f_\text{Nyq}$ of the input signal $\bmx=[x_1,\ldots,x_N]^T$ reduces to \revision{a uniform} sub-sampling rate $f_s$ of the measurements $\bmy=[y_1,\ldots,y_M]^T$ such that $N T_\text{Nyq}=M T_s$.
For \emph{uniform} sub-sampling at rate $f_s$ (i.e., we do \emph{not} perform NUS of the wavelet samples), the compression ratio $N/M$ is proportional to the sub-sampling ratio $\kappa=f_\text{Nyq}/f_s$.
\revision{If we randomly select a subset of samples of the sample stream (in addition to uniform sub-sampling), then we can further lower the (average) sampling rate, effectively implementing CS.}
\revision{The sampling diversity of NUWBS comes from the wavelet parameters settings, namely the width parameter $\tau$ and the central frequency $f^c$.}

\revision{While the architecture depicted in \figref{fig:implementation} is purely serial, one can deploy multiple parallel branches to (i) further increase the diversity of the CS acquisition stage, (ii) reduce the ADC rate by interleaved processing, or (iii) sense multiple sub-bands.}
In addition, a multi-branch architecture can simplify the circuitry for each branch by fixing the center frequency, pulse width, delay, or pulse rate per branch. 
\revision{For such an architecture, each branch performs wavelet bandpass sampling at a given scale with fixed bandwidth and center frequency.}

\subsection{Idealistic CWT Bandpass Sampling versus Realistic Serial Wavelet Bandpass Sampling}
This section discusses the commonalities and differences between idealistic CWT bandpass sampling  and the serial wavelet bandpass sampling architecture shown in \figref{fig:implementation}.

\subsubsection{Analysis of CWT Bandpass Sampling}
\label{sec:CWTbandpass}
The wavelet coefficient~$\setW x_{s,\delta}$ of the signal $x(t)$ at a scale $s$ and time shift $\delta$ is defined in \eqref{eq:waveletcoefficient}. Assume that the time-shift parameter~$\delta$ is continuous so that CWT is a continuous function in $\delta$. Then, the scalar product in \eqref{eq:timerelation} can be rewritten using the convolution operator~$*$ as follows~\cite{mallat1999wavelet}:
\begin{align} \label{eq:timerelation}
\setW x_s(\delta) = (x * \widetilde{\psi}_s)(\delta).
\end{align}
Here, $\widetilde{\psi}_s(u)=\frac{1}{\sqrt{s}}\psi^*(\frac{-u}{s})$. We can now compute the Fourier transform $\setF$ in the time-shift parameter $\delta$ to obtain 
\begin{align} \label{eq:fourierrelation}
\setF\{ \setW x_s(\delta)  \}  = X(f)  \widetilde{\Psi}_s(f),
\end{align}
\revision{where $\widetilde{\Psi}_s(f)$ is the Fourier transform of the wavelet $\widetilde{\psi}_s(u)$ given by 
\begin{align} \label{eq:fourierwavelet}
\widetilde{\Psi}_s(f)=\setF\big\{\widetilde{\psi}_s(u)\big\}=\sqrt{s}\Psi^*(-sf)
\end{align}
and $\Psi(f)$ is the Fourier transform of the mother wavelet~$\psi(t)$.
From \eqref{eq:fourierrelation}, we see that the CWT $\setW x_s(\delta)$ is equivalent to filtering the input signal $X(f)$ with the transfer function~$H_\text{CWT}(f)=\widetilde{\Psi}_s(f)$.}
We can now analyze the result of bandpass sampling applied to the function $\setW x_s(\delta)$. To this end, we assume a sampling rate $f_s$ well-below the Nyquist bandwidth of the input signal $x(t)$ and below the bandwidth of the mother wavelet, i.e., $f_s \le \textit{BW}_p   \ll f_\text{Nyq}$. 

We have the following discrete-time output signal 
\begin{align*}
y[t=nT_s] = \sum^{n=+\infty}_{n=-\infty} \setW x_s(n T_s) \delta(t-nT_s)
\end{align*}
sampled at $f_s=1/T_s$. 
The output signal $y[t=nT_s]$ corresponds to the bandpass sampled version of signal $x(t)$ after filtering it with the transfer function $H_\text{CWT}(f)=\widetilde{\Psi}_s(f)$. 
In contrast to classical bandpass sampling, the initial CWT extracts a particular frequency band defined by the center frequency~$f^c$ and the bandwidth $\textit{BW}_p$ of the wavelets. In words, CWT bandpass sampling is an effective combination of filtering and mixing via bandpass sampling. 
It is important to realize that this scheme requires access to the continuous-time CWT of the signal prior to sub-sampling. In practice, however, we do not have access to the CWT---instead, we have to make use of the wavelet sampling architecture shown in \figref{fig:implementation}.

Performing a CWT in hardware is infeasible and would require an excessively large number of branches, i.e., a dedicated branch per time shift~$\delta$ or convolution result every~$T_\text{Nyq}$ second as the CWT atoms have infinite support.
In contrast, the architecture proposed in \figref{fig:implementation} performs \revision{a convolution} of the input signal with the atom $\widetilde{\psi}_s(\delta)$ every $T_s$ second (instead of $T_\text{Nyq}$) in a serial manner.
While both approaches are similar, there are important differences in the filtering capabilities. To this end, we investigate the out-of-support $\Sigma$  (out-of-band interference) rejection performance for serial wavelet bandpass sampling that can be implemented (cf.~\figref{fig:implementation})  and  the idealistic CWT bandpass sampling approach.

\subsubsection{Analysis of Serial Wavelet Bandpass Sampling}
\begin{figure}[t] % figures should generally be on top of the page; not within the text
\centering
\includegraphics[width=0.95\columnwidth]{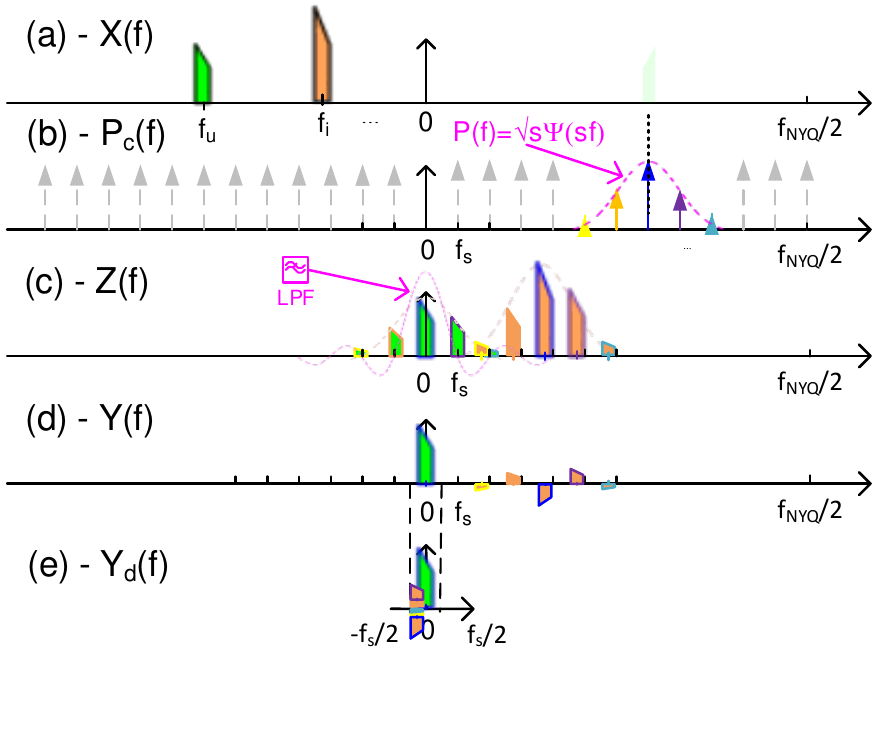}
\vspace{-1.1cm}
\caption{\revision{Illustration of the signal (useful $f_u$ and interferer $f_i$) spectrum evolution along the serial wavelet bandpass sampling; The wavelet sampling rate is $f_s$ with constant wavelet parameters settings $(\tau,f^c)$. The input signal spectrum $X(f)$ is convolved with weighted Dirac comb, filtered by a sinc low-pass filter, and decimated at rate~$f_s$.}}
\label{fig:pulsesubsampling}
\end{figure}

Consider the case  in which both the wavelet center  frequency and bandwidth remains constant for the entire wavelet comb. This is the case of the Gabor frame projection reported on a single branch in \figref{fig:selection}.  \revision{We will use \figref{fig:pulsesubsampling}, which illustrates the spectrum representation, to assist our discussion. The input signal $x(t)$ in \figref{fig:pulsesubsampling}(a) is first multiplied (mixed) with a wavelet comb $p_c(t)$ shown in \figref{fig:pulsesubsampling}(b). The mixing result $z(t)$ can be expressed as follows:}
\begin{align*}
z(t) = x(t)  p_c(t) = x(t)  \sum^{n=+\infty}_{n=-\infty} p(t-nT_s),
\end{align*}
where $p(t)$ is the considered wavelet.  
The Fourier transform of the signal $z(t)$ shown in \figref{fig:pulsesubsampling}(c) is given by 
\begin{align} \label{eq:mixspectrum}
Z(f) = f_s X(f) *  P(f)\sum^{k=+\infty}_{k=-\infty}  \delta(f-k f_s),
\end{align}
which reveals that the spectrum of the mixed signal $Z(f)$ is the convolution between the Fourier transform of the input signal $X(f)$ (cf.~\figref{fig:pulsesubsampling}(a)) and a Dirac comb weighted by the Fourier transform of the wavelet~$P(f)$. According to \eqref{eq:waveletfrequency} the Gaussian envelope of Gabor atoms or C-Morlet wavelets show exponentially fast decay, which implies that the infinite sum can effectively be reduced to a small number of Dirac delta functions shown in \figref{fig:pulsesubsampling}(b). 
\revision{Furthermore, we see that NUWBS effectively reduces noise folding by pre-filtering the spectrum with the pulse~$P(f)$ prior to band-pass sampling; this is contrast to conventional band-pass sampling in which noise from the entire Nyquist bandwidth  folds into each sample~\cite{vaughan1991signal}.}

In order to match this approach with the bandpass CWT approach discussed in \secref{sec:CWTbandpass}, we can see that $p(t)$ corresponds to the wavelet atom at the scale $s$ and time shift $\delta=0$ with Fourier transform
\begin{align} \label{eq:pulse}
P(f) = \setF\!\left\{\frac{1}{\sqrt{s}}\Psi\!\left(\frac{t}{s}\right)\right\}= \sqrt{s}\Psi(sf).
\end{align}
By comparing \eqref{eq:fourierwavelet} with \eqref{eq:pulse}, we see that one is the complex conjugate of the other.
In the architecture shown in \figref{fig:implementation}, the mixing product $z(t)$ is low-pass filtered. A typical filter that can be implemented corresponds to an integration over a rectangular window of duration $T_s$. The frequency-domain representation of this integrator corresponds to the  cardinal sine (sinc) function. \revision{Hence, the Fourier transform $Y(f)$ of the filtered and mixed signal $Z(f)$ shown in \figref{fig:pulsesubsampling}(d) is}  
\begin{align} \label{eq:filteredsig}
Y(f) = Z(f) T_s \mathrm{sinc}(T_s f),
\end{align}
where we define $\mathrm{sinc}(u)=\sin(\pi u)/(\pi u)$.
In the architecture shown in \figref{fig:implementation}, the signal $y(t)$ is finally decimated by a factor $\kappa$ such that $\kappa=f_\text{Nyq}/f_s$, i.e., the entire Nyquist band is folded into the frequency range $[-f_s/2,f_s/2]$.  Hence, the sample stream of the decimated signal is 
\begin{align*}
y_d[nT_s] = y[n\kappa \Delta t] \quad \text{with} \quad \Delta t=1/f_\text{Nyq}
\end{align*}
and the Fourier transform of the discrete signal $y_d[nT_s]$ shown in \figref{fig:pulsesubsampling}(e) is given by
\begin{align} \label{eq:decimatedsig}
Y_d[e^{2j\pi f}] = \frac{1}{\kappa} \sum_{r=0}^{\kappa-1} Y\!\left(e^{2j\frac{f-r}{\kappa}}\right).
\end{align}
According to this equation, the serial wavelet bandpass sampling method collapses all the sinc-filtered and Gaussian weighted convolution products into the band $[-f_s/2,f_s/2]$. As illustrated in \figref{fig:pulsesubsampling}(e), because of the sub-sampling process, the output frequency location is folded to $\{f_i/f_s\}f_s$ with $\{f_i/f_s\}$ the fractional part between the interference frequency $f_i$ and the wavelet repetition rate, equal in our case to the output sampling frequency $f_s$. The equivalent filtering effect is given by 
\begin{align}
H_\text{WBS}(f) & = \sum_{k=-\kappa/2}^{\kappa/2-1} \mathrm{sinc}(T_s(f-kf_s)) P(k f_s) \notag \\
 & = \sum_{k=-\kappa/2}^{\kappa/2-1} \mathrm{sinc}(T_s(f-kf_s)) e^{-(\pi\tau kf_s)^2}.   \label{eq:Hps}
\end{align}
\revision{The expression in \eqref{eq:Hps} highlights the out-of-band rejection capabilities of the proposed (realistic) serial wavelet bandpass sampling approach in comparison with the (idealistic) CWT bandpass method computed in \eqref{eq:fourierrelation}. We emphasize that the major differences between the serial wavelet bandpass sampling approach and the CWT baseband sampling comes from the fact that the equivalent filter transfer function differs from a mixture of sinc-shaped for the former (see Eq.~\ref{eq:Hps}) to a Gaussian shape (with infinite support) for the latter (see Eq.~\ref{eq:fourierwavelet}).}

\subsubsection{Simulation Results}
We now validate the serial wavelet bandpass sampling scheme and more specifically evaluate the out-of-band rejection capabilities. As illustrated in \figref{fig:pulsesubsampling}(a), the input signal $x(t)$ is complex-valued and builds upon a useful signal located at $f_u$ within the band of interest (we assume $f_s$ is a sub-multiple of $f_c$) and an out-of-band interference signal at $f_i$ located $\Delta f_i$ apart from our signal of interest. The signal $x(t)$ is sub-sampled by a uniform wavelet comb at rate $f_s=1/(4\tau)$. We consider a sampling rate of $f_s=1$\,GHz. The  wavelet parameters, such as width parameter $\tau$ and central frequency~$f^c$, remain constant \revision{over the frame while the time shift is adjusting to the sampling position}.

As shown in \eqref{eq:mixspectrum} and illustrated in \figref{fig:pulsesubsampling}(b), serial wavelet bandpass sampling is, in the frequency domain, equivalent to a Dirac comb whose amplitude is weighted by the wavelet (or pulse) envelope $P(f)$.  Since $f_s<\textit{BW}_p$ (because $1/f_s=T_s=4\tau$) temporal overlapping among wavelets is avoided, several Dirac functions are included within the pulse envelope centered on carrier frequency $f_c$. As illustrated in \figref{fig:pulsesubsampling}(c), each convolution down-converts the useful signal to the origin  and out-of-band interferences into baseband. Then, the integration over a time period $T_s$    low-pass filters the signals that are close to baseband (see \figref{fig:pulsesubsampling}(d)).

\begin{figure}[tp]
\centering
\includegraphics[width=0.95\columnwidth]{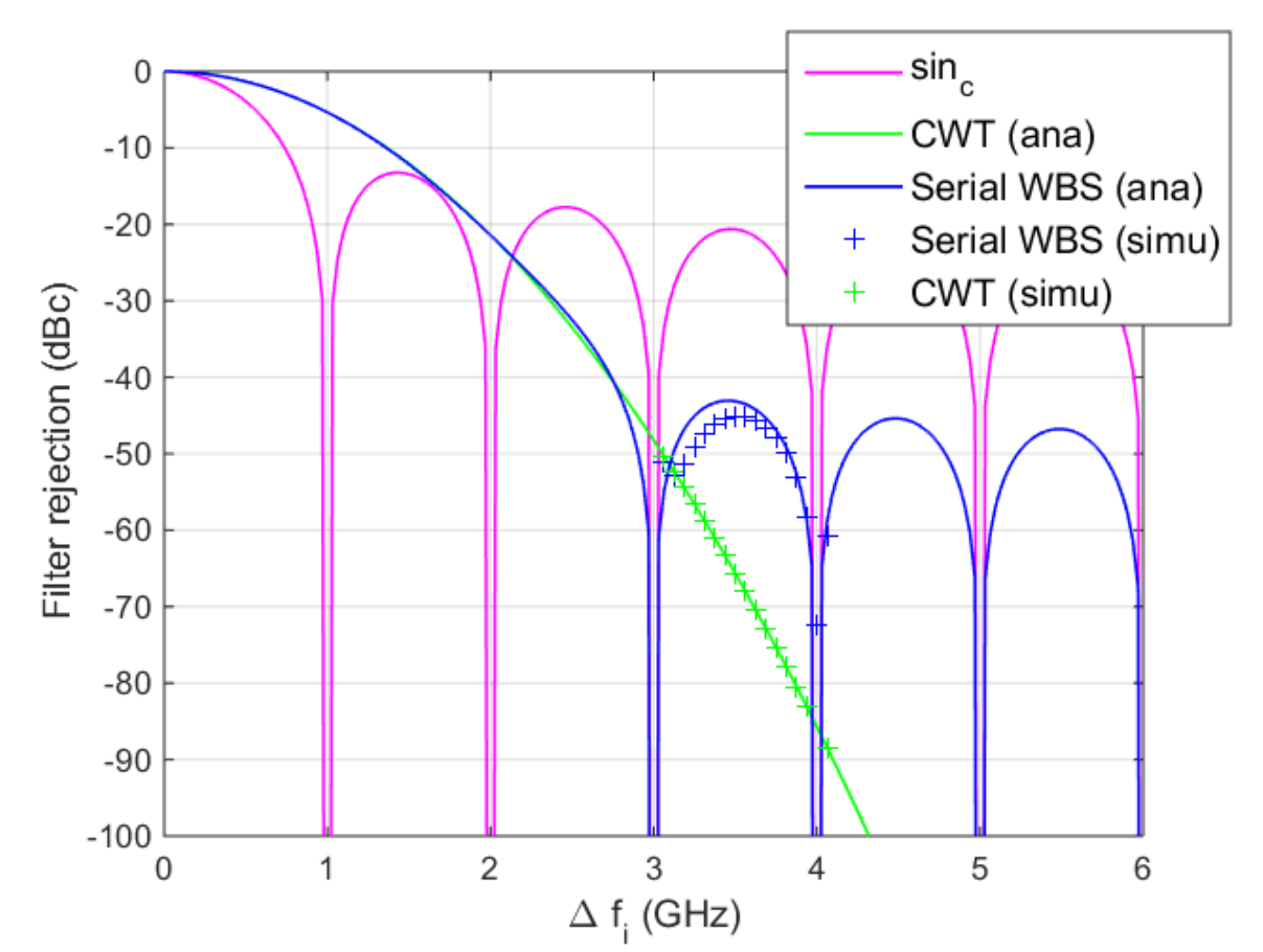}
\vspace{-0.1cm}
\caption{\revision{Comparison between out-of-band interference rejection between the CWT and wavelet bandpass sampling in the case of wavelet sampling rate equal to four times the wavelet bandwidth (i.e., $T_s=4\tau=1$\,ns).}}
\label{fig:realvstheorypulse}
\end{figure}

\revision{\figref{fig:realvstheorypulse} summarizes the out-of-band rejection characteristics of the serial wavelet bandpass sampling approach $H_\text{WBS}(f)$ and provides a comparison with the idealistic CWT~$H_\text{CWT}(f)$. This analysis quantifies the out-of-band alias rejection capability of NUWBS. Our analytical expressions from~\eqref{eq:Hps} and \eqref{eq:fourierwavelet} are shown with continuous lines;  simulation results are indicated with plus ($+$) markers, respectively, in blue for wavelet bandpass sampling and green for the idealistic CWT method.}

Evidently, our simulations coincide with the theoretical results in~\eqref{eq:Hps} and \eqref{eq:fourierwavelet}. We also observe that the filter characteristics  of serial wavelet bandpass sampling  is in-between the ideal equivalent Gaussian filter and the standard sinc filter associated to the $T_s$ rectangular windows integration. As a result, we achieve a rejection of   $50$\,dBc, which remains to be lower than the idealistic CWT rejection but (i) with more than  $23$\,dB improvement with respect to the standard sinc filter and (ii) can, as shown next, be implemented in hardware. 

\subsection{Wavelet Generator Circuit}

\begin{figure}[tp]
\centering
\includegraphics[width=0.95\columnwidth]{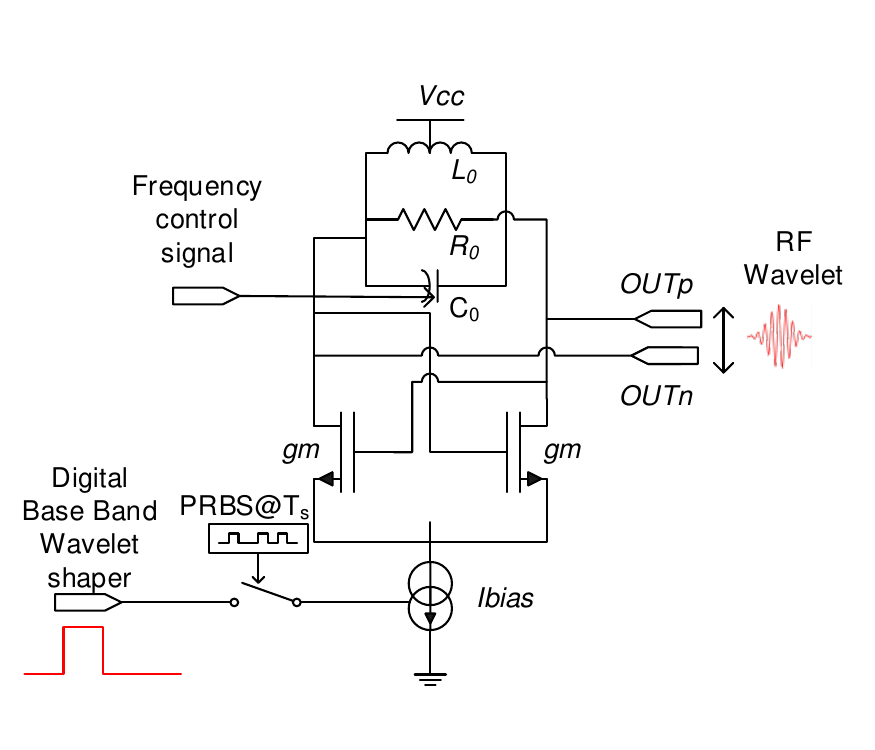}
\vspace{-0.5cm}
\caption{Circuit schematic of a wavelet pulse generator with variable bandwidth and central frequency capabilities. The circuit acts as a Voltage Control Oscillator (VCO) switched on according to a sub-Nyquist PRBS sequence.}
\label{fig:pulsegenerator}
\end{figure}

The key missing piece of the proposed NUWS and NUWBS approach is the tunable wavelet generator circuit.
For RF applications, wavelet generation in the time domain can be realized by leveraging extensive prior work in the field of ultra-wideband (UWB) impulse technology  \cite{apsel2013design}.
For instance, in our previous work  \cite{pelissier2011a112}, we have demonstrated a circuit for low-power pulse generation at $8$\,GHz with variable pulse repetition rate. Here, we suggest to adapt the design in \cite{pelissier2011a112}, for tunable and wideband wavelet generation. 
\figref{fig:pulsegenerator} shows a corresponding circuit diagram. 
The core of the oscillator relies on  a cross-coupled NMOS pair loaded by an RLC resonator highlighted, which is commonly used for voltage control oscillator  (VCO) circuits. 
Wavelets are generated across the LC tank at RF frequency as soon as the bias current $I_\text{bias}$ is applied to the cross-coupled pair. 

\begin{figure}[tp]
\centering
\vspace{-0.45cm}
\includegraphics[width=0.95\columnwidth]{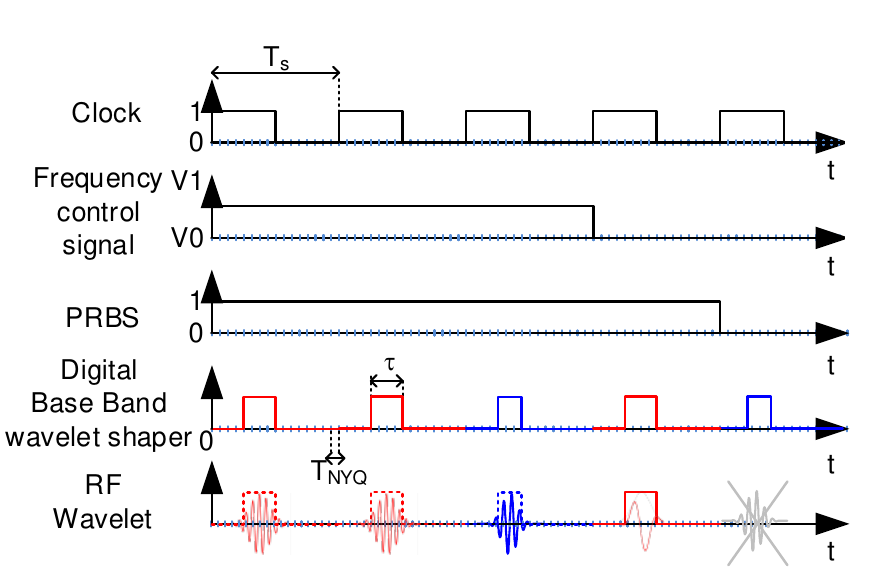}
\vspace{-0.2cm}
\caption{Signal \revision{chronogram} involved in the control of the circuit schematic shown in \figref{fig:pulsegenerator}: The frequency control signal, the digital base band pulse shaper, and the clock and PRBS signals are running at sub-Nyquist rates.}
\label{fig:chronogram}
\end{figure}

\revision{\figref{fig:chronogram} shows a typical chronogram of the proposed wavelet generation circuit. The bias current duration is adjusted by a digital base-band pulse shaper to enable variable bandwidth. A clock signal running at rate $f_s$ is combined with a pseudo-random bit-sequence (PRBS) running at the low sub-sampling rate $f_s$ in order to switch  the biasing source on and off. As a result a non-uniform pulse pattern is generated tailored to the NUWBS solution. Finally, a variable voltage applied to the varactor $C_0$ in \figref{fig:pulsegenerator} enables us to tune the center frequency to the RF sub-band of interest.}


\begin{figure}[tp]
\centering
\includegraphics[width=0.95\columnwidth]{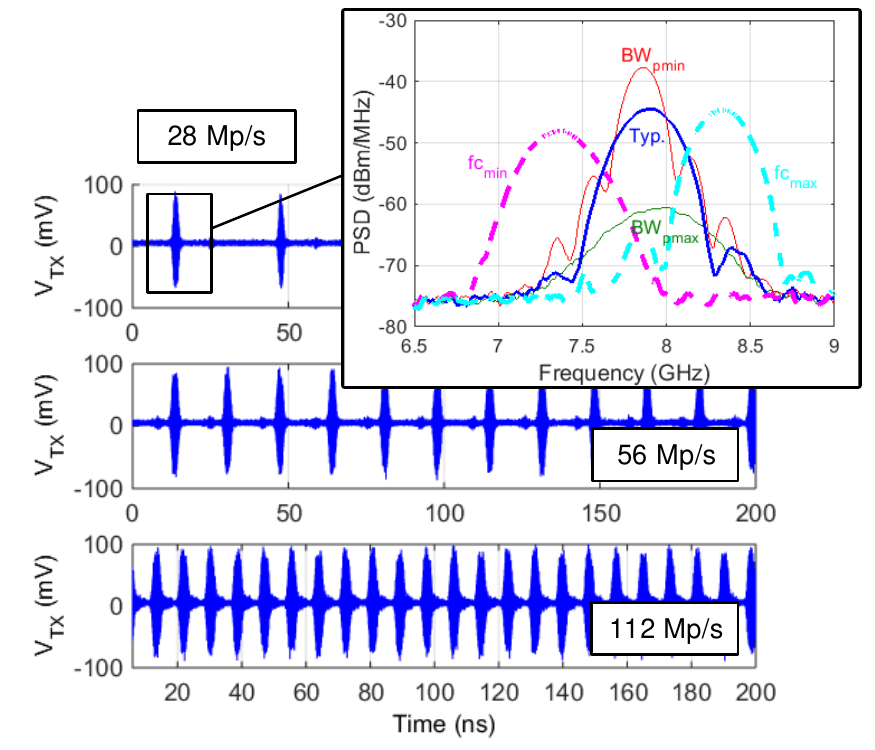}
\vspace{-0.2cm}
\caption{ASIC measurement results: Illustration of variable wavelet rate generation (28/56/112\,Mp/s); Zoom in on a single wavelet spectrum illustrating the wavelet central frequency and bandwidth tuning capabilities.}
\label{fig:realpulse}
\end{figure}

\revision{In order to validate the wavelet generator circuitry for RF applications up to 8\,GHz, physical measurements have been performed on an ASIC fabricated in a $130$\,nm CMOS technology.} \figref{fig:realpulse} shows the power spectral density (PSD) of the wavelet depending on the bandwidth or central frequency. 
\revision{Measurements are provided at $28.125$\,Mp/s, $56.25$\,Mp/s, and $112.5$\,Mp/s with amplitude up to $160$\,mV for $50$\,$\Omega$ impedance. The Tektronix TDS6124C high speed scope is set to $50$\,$\Omega$ impedance to avoid any reflections with lab equipment that could alter the wavelet waveform. A high timing resolution mode with digital interpolation between the $25$\,ps real samples is selected to provide a $5$\,ps timing resolution.}
The $-10$\,dB wavelet bandwidth is tunable from $300$\,MHz to $1$\,GHz and the central frequency range from $7.3$\,GHz to $8.5$\,GHz. In addition to being flexible, the wavelet generation is power efficient, i.e., only requires $60$\,pJ/pulse, and remains switched off in between two successive wavelet generation phases (i.e., this is duty-cycled solution).
Our ASIC measurements results demonstrate a feasible wavelet function generator with a broad range of  tuning capabilities in terms of central frequency, bandwidth, and repetition rate operating in the range of RF frequencies. 
\revision{These results pave the way for a complete NUWS/NUWBS integration including signal mixing and sampling stage.}

% ================================================================================
% ================================================================================
% ================================================================================

\section{Conclusions}
\label{sec:conclusions}
We have proposed a novel analog-to-information (A2I) conversion method for compressive-sensing (CS)-based RF feature extraction. 
Our approach, referred to as non-uniform wavelet sampling (NUWS), combines wavelet preprocessing with non-uniform sampling (NUS), which mitigates the main issues of existing analog-to-information (A2I) architectures, such as out-of-band noise, interference, aliasing, and flexibility. In addition, NUWS avoids circuitry that must adhere to Nyquist rate bandwidths. From an RF feature extraction standpoint, NUWS can be adapted to the signals of interest by tuning their duration, center frequency, and time instant per acquired wavelet sample.

For multiband RF signals, we have developed a specialized variant of NUWS called non-uniform wavelet bandpass sampling (NUWBS). For this method, we have discussed a wavelet selection strategy that enables adaptation to the a-priori knowledge of the sub-bands of interest.
Using simulation results, we have shown that NUWBS achieves near-optimal sample complexity already for relatively small dimensions, i.e., NUWBS approaches the theoretical phase transition of $\ell_1$-norm-based sparse signal recovery with Gaussian measurement ensembles. 
We have furthermore analyzed the rejection rate of NUWBS against out-of-band interferers. To demonstrate the practical feasibility of our A2I feature extractor, we have proposed a suitable wavelet generation circuit that enables the generation of tunable wavelet pulses  in the GHz regime.

The proposed NUWS and NUWBS methods are promising strategies for A2I converter architectures that overcome the traditional limitations of existing solutions in power and cost limited applications. 
Our solutions find potential broad use in a variety of RF receivers targeting spectrum awareness or assisting conventional RF chains with tuning parameters. Both of these advantages render our solutions useful for the Internet of Things, for which power and cost efficiency and RF feature extraction are of utmost importance.

\revision{There are many avenues for future work. The design of a complete NUWS/NUWBS-based RF feature extractor ASIC is ongoing work. A theoretical analysis of the recovery properties for NUWS/ NUWBS is a challenging open research problem. 
Finally, a detailed exploration of other applications that may benefit of NUWS/NUWBS and are in need of low power and low cost feature extraction is left for future work.}

\section*{Acknowledgments}
The work of M.~Pelissier was supported by the  Enhanced Eurotalents fellowships program \& Carnot Institut. \revision{The work of C.~Studer was supported by Xilinx, Inc.\ and by the  US National Science Foundation under grants CCF-1535897,  ECCS-1408006, and CAREER CCF-1652065.} The authors thank O.~Casta\~neda for his help with the manuscript preparation.

% ================================================================================

\balance

%\bibliographystyle{IEEEtran} 
%\bibliography{cites}


\balance

\end{document}